\documentclass[twocolumn]{aastex63}

\usepackage{amsmath, bm, empheq, mathrsfs,  cancel}
\usepackage[counterclockwise]{rotating}
\usepackage{natbib}

\newcommand{\pderiv}[2]{\frac{\partial#1}{\partial#2}}
\newcommand{\matderiv}[1]{\frac{D#1}{Dt}}
\newcommand{\pderivline}[2]{\partial#1/\partial#2}

\newcommand{\av}[1]{\left\langle#1\right\rangle}
\newcommand{\avsph}[1]{\left\langle#1\right\rangle_{\rm{sph}}}
\newcommand{\avspht}[1]{\left\langle#1\right\rangle_{ {\rm sph}, t}}
\newcommand{\avt}[1]{\left\langle#1\right\rangle_{t}}
\newcommand{\avphi}[1]{\left\langle#1\right\rangle_{\phi}}
\newcommand{\avphit}[1]{\left\langle#1\right\rangle_{\phi,t}}

\newcommand{\avaltsph}[1]{\langle#1\rangle_{\rm{sph}}}

\newcommand{\sn}[2]{#1\times10^{#2}}

\newcommand{\definealt}{\equiv}

\newcommand{\five}{\ \ \ \ \ }
\newcommand{\orr}{\text{or}\five }
\newcommand{\andd}{\text{and}\five }
\newcommand{\where}{\text{where}\five }

\newcommand{\curl}{\nabla\times}
\newcommand{\Div}{\nabla\cdot}

\newcommand{\dotgrad}{\cdot\nabla}
\newcommand{\ugrad}{\bm{u}\dotgrad}

\newcommand{\e}{\hat{\bm{e}}}
\newcommand{\er}{\e_r}
\newcommand{\et}{\e_\theta}

\newcommand{\ez}{\e_z}

\newcommand{\ofr}{(r)}

\newcommand{\rhoref}{\overline{\rho}}
\newcommand{\tmpref}{\overline{T}}
\newcommand{\prsref}{\overline{P}}
\newcommand{\sref}{\overline{S}}
\newcommand{\gref}{\overline{g}}
\newcommand{\nuref}{\overline{\nu}}
\newcommand{\kapparef}{\overline{\kappa}}
\newcommand{\etaref}{\overline{\eta}}
\newcommand{\nsqref}{\overline{N^2}}
\newcommand{\qref}{\overline{Q}}
\newcommand{\cref}{\overline{C}}
\newcommand{\fluxnr}{\overline{F_{\rm nr}}}
\newcommand{\fluxnrtilde}{\widetilde{F_{\rm nr}}}

\newcommand{\dsdr}{\frac{d\overline{S}}{dr}}

\newcommand{\dsdrline}{d\overline{S}/dr}

\newcommand{\cv}{c_{\rm{v}}}
\newcommand{\cp}{c_{\rm{p}}}

\newcommand{\prspert}{P}
\newcommand{\spert}{S}

\newcommand{\vecu}{\bm{u}}
\newcommand{\vecb}{\bm{B}}
\newcommand{\vecom}{\bm{\omega}}

\newcommand{\upol}{\vecu_{\rm{pol}}}
\newcommand{\bpol}{\vecb_{\rm{pol}}}

\newcommand{\urad}{u_r}

\newcommand{\inn}{_{\rm{in}}}
\newcommand{\out}{_{\rm{out}}}

\newcommand{\rms}{_{\rm{rms}}}
\newcommand{\const}{_{\rm{const}}}

\newcommand{\lmax}{{\ell_{\rm{max}}}}

\newcommand{\rsun}{R_\odot}

\newcommand{\lsun}{L_\odot}
\newcommand{\omsun}{\Omega_\odot}

\newcommand{\msun}{M_\odot}

\newcommand{\taurs}{\tau_{\rm{rs}}}
\newcommand{\taurad}{\tau_{\rm{rad}}}
\newcommand{\taums}{\tau_{\rm{ms}}}
\newcommand{\taumm}{\tau_{\rm{mm}}}
\newcommand{\taumc}{\tau_{\rm{mc}}}
\newcommand{\tauv}{\tau_{\rm{v}}}
\newcommand{\taumag}{\tau_{\rm{mag}}}

\newcommand{\pes}{{P_{\rm{ES}}}}
\newcommand{\pessun}{{P_{ {\rm ES}, \odot}}}
\newcommand{\pnu}{{P_{\nu}}}
\newcommand{\pkappa}{{P_{\kappa}}}
\newcommand{\peta}{{P_{\eta}}}
\newcommand{\prot}{{P_{\rm{rot}}}}

\newcommand{\tequil}{{t_{\rm{eq}}}}
\newcommand{\tmax}{{t_{\rm{max}}}}
\newcommand{\pcyc}{{P_{\rm{cyc}}}}
\newcommand{\omcyc}{{\omega_{\rm{cyc}}}}

\newcommand{\ra}{{\rm{Ra}}}

\newcommand{\raf}{\ra_{\rm{F}}}
\newcommand{\rafmod}{\raf^*}

\newcommand{\pr}{{\rm{Pr}}}
\newcommand{\prm}{{\rm{Pr_m}}}

\newcommand{\di}{{\rm{Di}}}

\newcommand{\ek}{{\rm{Ek}}}
\newcommand{\ta}{{\rm{Ta}}}

\newcommand{\ro}{{\rm{Ro}}}

\newcommand{\roc}{{\rm{Ro_c}}}

\newcommand{\re}{{\rm{Re}}}
\newcommand{\rem}{{\rm{Re_m}}}

\newcommand{\gram}{{\rm{g}}}
\newcommand{\cm}{{\rm{cm}}}

\newcommand{\gauss}{{\rm{G}}}
\newcommand{\kelv}{{\rm{K}}}
\newcommand{\unitent}{{\rm{erg\ g^{-1}\ K^{-1}}}}
\newcommand{\uniten}{\rm{erg}\ \cm^{-3}}
\newcommand{\unitprs}{\rm{dyn}\ \cm^{-2}}
\newcommand{\unitrho}{\gram\ \cm^{-3}}
\newcommand{\stoke}{\rm{cm^2\ s^{-1}}}

\newcommand{\rayleigh}{\texttt{Rayleigh}}

\newcommand{\eulag}{\texttt{EULAG}}

\newcommand{\ash}{\texttt{ASH}}

\defcitealias{Matilsky2022}{Paper I}

\newcommand{\cz}{_{\rm{CZ}}}
\newcommand{\rz}{_{\rm{RZ}}}
\newcommand{\full}{_{\rm{full}}}
\newcommand{\dimm}{_{\rm{dim}}}

\newcommand{\omnyq}{{\omega_{\rm{nyq}}}}
\newcommand{\rtsun}{r_{t,\odot}}

\newcommand{\bpolperm}{\vecb_{\rm pol,perm}}
\newcommand{\newtext}[1]{#1}

\received{17 November 2023}
\revised{22 December 2023}
\accepted{23 December 2023}
\submitjournal{ApJ}

\shorttitle{Confinement of the Solar Tachocline by a Non-Axisymmetric Dynamo}
\shortauthors{Matilsky et al.}

\begin{document}
	
	\title{Confinement of the Solar Tachocline by a Non-Axisymmetric Dynamo}

	\correspondingauthor{Loren I. Matilsky}
	\email{loren.matilsky@gmail.com}	
	\author[0000-0001-9001-6118]{Loren I. Matilsky}\thanks{NSF Astronomy and Astrophysics Postdoctoral Fellow}	
	\affiliation{Department of Applied Mathematics,
	Baskin School of Engineering,
	University of California, 
	Santa Cruz, CA 96064-1077, USA}

	\author[0000-0003-4350-5183]{Nicholas H. Brummell}
	\affiliation{Department of Applied Mathematics,
	Baskin School of Engineering,
	University of California, 
	Santa Cruz, CA 96064-1077, USA}	

	\author[0000-0001-7612-6628]{Bradley W. Hindman}
	\affiliation{Department of Applied Mathematics,
	University of Colorado,
	Boulder, CO 80309-0526, USA}
	\affiliation{JILA \& Department of Astrophysical and Planetary Sciences,
	University of Colorado,
	Boulder, CO 80309-0440, USA}
	
	\author[0000-0002-3125-4463]{Juri Toomre}
	\affiliation{JILA \& Department of Astrophysical and Planetary Sciences,
		University of Colorado,
		Boulder, CO 80309-0440, USA}
	
	\begin{abstract}
	We recently presented the first 3D numerical simulation of the solar interior for which tachocline confinement was achieved by a dynamo-generated magnetic field. In this followup study, we analyze the degree of confinement as the magnetic field strength changes (controlled by varying the magnetic Prandtl number) in a coupled radiative zone (RZ) and convection zone (CZ) system. We broadly find three solution regimes, corresponding to weak, medium, and strong dynamo magnetic field strengths. In the weak-field regime, the large-scale magnetic field is mostly axisymmetric with regular, periodic polarity reversals (reminiscent of the observed solar cycle), but fails to create a confined tachocline. In the strong-field regime, the large-scale field is mostly non-axisymmetric with irregular, quasi-periodic polarity reversals, and creates a confined tachocline. In the medium-field regime, the large-scale field resembles a strong-field dynamo for extended intervals, but intermittently weakens to allow temporary epochs of strong differential rotation. In all regimes, the amplitude of poloidal field strength in the RZ is very well explained by skin-depth arguments, wherein the oscillating field that gives rise to the skin depth (in the medium- and strong-field cases) is a non-axisymmetric field structure rotating with respect to the RZ. These simulations \newtext{suggest a new picture of solar tachocline confinement by the dynamo, in which non-axisymmetric, very long-lived (effectively permanent) field structures rotating with respect to the RZ play the primary role, instead of the regularly reversing axisymmetic field associated with the 22-year cycle.}
	\end{abstract}
	
	

	\section{The Solar Tachocline} \label{sec:intro}
 The solar tachocline is a region of primarily radial shear at the base of the solar convection zone (CZ), where strong latitudinal differential rotation transitions to nearly solid-body rotation in the underlying radiative zone (RZ). The tachocline is observed helioseismically to be centered at $\rtsun\approx0.69\rsun$ (which roughly coincides with the base of the CZ) and to have a thickness of $\Gamma_\odot\lesssim0.05\rsun$ ($\Gamma_\odot$ is too small to be helioseismically resolved, implying that it has an upper bound roughly equal to the helioseismic inversion kernel width; e.g., \citealt{Howe2009}). Some measurements estimate a wider tachocline ($\Gamma_\odot\lesssim0.10\rsun$; e.g., \citealt{Kosovichev1996, Wilson1996}) or a narrower tachocline ($\Gamma_\odot\lesssim0.02\rsun$; e.g., \citealt{Elliott1997, Basu2003}). 
 
 Regardless of the true tachocline thickness, even the most liberal estimates for $\Gamma_\odot$ pose a major dynamical problem for solar physics. It is hypothesized \citep{Spiegel1992} that the CZ's differential rotation should spread into the RZ by a process similar to circulation ``burrowing" in rotating stably stratified shear flows (e.g., \citealt{Clark1973, Haynes1991}), thus widening the tachocline. A shear flow in a rotating system (i.e., differential rotation) is usually accompanied by a horizontal temperature gradient due to thermal wind balance (e.g., \citealt{Aurnou2011, Matilsky2023}). This gradient tends to spread (burrow) further into the stable layer via thermal conduction, carrying with it the circulation and differential rotation associated with the thermal wind. In the Sun, the dominant thermal diffusion is radiative, and \citet{Spiegel1992} showed that burrowing (now referred to as ``radiative spread") should have increased $\Gamma_\odot$ to $\sim$$0.4\rsun$ by the current age of the Sun. 
 
 \citet{Spiegel1992}'s original argument that the solar tachocline should radiatively spread assumes axisymmetry and linearized fluid equations. Under those conditions, radiative spread occurs ``hyperdiffusively" (governed by $\nabla^4$ instead of $\nabla^2$) on the solar Eddington-Sweet time $\pessun$. In the \newtext{hyperdiffusive} case, $\Gamma(t)/\Gamma_\odot\sim(t/\pessun)^{1/4}$, where $\Gamma(t)$ is the time-dependent tachocline thickness and $t$ is the time since initial confinement [i.e., $\Gamma(0)=\Gamma_\odot$]. Since the Eddington-Sweet time is so long for the Sun ($\pessun\approx600$ Gyr; see Table \ref{tab:solaranalog}), this hyperdiffusive property is essential for the burrowing to be significant on time-scales as small as the solar age ($\sim$5 Gyr). Recent 3D fully nonlinear simulations have shown that circulation burrowing does indeed occur in more realistic settings (as long as the time-scales are properly ordered; see \citealt{Wood2012, Wood2018}). But whether realistic solar burrowing would be hyperdiffusive is still an open question and requires further investigation. 
 
If circulation burrowing is indeed significant for the Sun, it is obvious that there must be a confining (or ``rigidifying") torque in the RZ to keep $\Gamma_\odot$ under the helioseismically constrained upper bound. There are currently two dominant tachocline confinement scenarios that postulate the origin of this torque. The first, proposed by \citet{Spiegel1992}, is essentially hydrodynamic. It is supposed that hydrodynamic shear instabilities associated with the differential rotation create turbulence with predominantly horizontal motion, owing to the strong convectively stable stratification of the RZ. The Reynolds stresses from this horizontal turbulence then act like an enhanced horizontal viscosity, causing preferentially horizontal angular momentum transport, thereby eliminating any burrowing shear on the relatively fast time-scale of months to years. Hence, this scenario is often also called the ``fast confinement scenario" (e.g., \citealt{Gilman2000,Brun2017b}). 

However, similar horizontal turbulence in the Earth's stratosphere is theorized to be ``anti-diffusive," that is, transporting angular velocity up the rotation gradient instead of down it and driving the system away from solid-body rotation (e.g., \citealt{Starr1968, McIntyre1994}). In any event, angular momentum transport by stratified turbulence in a solar-like system is likely more complicated than simply ``diffusive or anti-diffusive." For example, \citet{Tobias2007} argue that horizontal turbulence in the presence of a weak toroidal magnetic field creates Maxwell stresses that nearly exactly cancel the Reynolds stresses, yielding zero net momentum transport. Finally, it remains unclear exactly how anisotropic stratified turbulent transport really is. For example, recent 3D direct numerical simulations \citep{Cope2020,Garaud2020} show that meanders of the streamwise flow (in a sufficiently turbulent regime) can vary on small vertical length-scales until secondary vertical shear instabilities (and associated vertical momentum transport) develop.
 
 \citet{Gough1998} proposed an alternative, magnetic confinement scenario. They argued that a weak (minimum $\sim$1 G) poloidal magnetic field in the RZ could resist the shearing motion of any imposed differential rotation via magnetic tension. This magnetic torque would be generated on the time-scale of radiative spread, namely some fraction of $\pessun$. Hence, \citealt{Gough1998}'s scenario is sometimes called the ``slow confinement scenario." Note that the fast confinement scenario is mostly hydrodynamic (with magnetism possibly playing a secondary role in modifying the primary baroclinic and shear instabilities), while the slow confinement scenario is fundamentally magnetic. 
 
 Finally, a ``fast magnetic confinement scenario" has been proposed and modeled in 1D (e.g., \citealt{ForgcsDajka2001,Barnabe2017}). Here, source of the RZ's confining poloidal field is the cycling solar dynamo (with the cycle period of $\sim$11 yr, i.e., fast compared to $\pessun$, but slow compared to time-scales associated with most hydrodynamic instabilities) diffusing downward to a skin depth. 
 
 \newtext{In prior global simulations of solar-like CZ--RZ systems, the chosen parameters have made radiative spread insignificant on the time-scales the simulations can be run. Nevertheless, significant \textit{viscous} spread occurs (see Section \ref{sec:torque}) and simulated tachoclines have been confined against this viscous spread through a variety of mechanisms. \citet{Browning2006} used combined mechanical and thermal forcing to explicitly impose a steady-state tachocline in a simulation using the {\ash} code. Further simulations using the {\ash} and {\rayleigh} codes---which are direct numerical simulation (DNS) codes---have implemented temporary, slowly spreading tachoclines through significantly lowered values of the viscosity in the RZ compared to the CZ (e.g., \citealt{Augustson2013,Brun2017,Bice2022}). Finally, the implicit large-eddy simulation (ILES) code {\eulag} ensures very small effective numerical viscosity in stable regions owing to the nature of the ILES time-stepping algorithm MPDATA \citep{Prusa2008}. On the time-scales for which {\eulag} simulations are run, both viscous and radiative spread are thus negligible and tachoclines that are effectively steady can occur (in both magnetic and purely hydrodynamic cases) without an explicit confinement mechanism being necessary (e.g., \citealt{Guerrero2013,Beaudoin2018}).}
 
We recently presented (\citealt{Matilsky2022}; hereafter \citetalias{Matilsky2022}; see also \citealt{Matilsky2021}), \newtext{the first 3D, spherical-shell simulation (in our case, a DNS) to achieve a steady-state tachocline that was self-consistently confined against explicit viscous spread. The source of the confinement was magnetic torque, which was in turn generated by} a non-axisymmetric, quasi-periodic dynamo. In the CZ, the magnetism was topologically similar to the ``partial wreaths" (longitudinally elongated bands of intense toroidal magnetism, with alternating polarity in longitude) identified in our prior CZ-only dynamos \citep{Matilsky2020a, Matilsky2020c}. We showed in \citet{Matilsky2020a} that the partial wreaths in the CZ-only case tended to form a  long-lasting magnetic structure that more or less rotated rigidly in a preferred frame. In the combined CZ--RZ tachocline systems considered in the current work, the partial wreaths rotate with respect to the RZ below. As far as the rigidly-rotating RZ is concerned, the partial wreaths above resemble a periodically reversing poloidal field and therefore the field diffusively imprints from the overshoot layer to a depth in the RZ consistent with the electromagnetic skin effect. 

 The main conclusion of the present paper is that the confinement mechanism identified in \citetalias{Matilsky2022} can be regarded as a more general version of the fast magnetic confinement scenario that stays robust in a wider parameter space (containing multiple cycling frequency components of the dynamo) and in a 3D geometry with a fully coupled CZ and RZ. Furthermore, a rotating, large-scale \textit{non-axisymmetric} poloidal field structure takes the place of the reversing \textit{axisymmetric} magnetism (``full wreaths") typically invoked in connection with the observed solar cycle, or magnetic butterfly diagram. Our evidence consists of a family of solutions related to the one from \citetalias{Matilsky2022}, but with a range of magnetic Prandtl numbers $\prm$. One key effect of varying $\prm$ (while keeping the other control parameters fixed) is to achieve a range of magnetic field strengths in the saturated dynamo state, while keeping other key diagnostic parameters (like the Reynolds and Rossby numbers) relatively unchanged. 

The rest of this paper is structured as follows. In Section \ref{sec:exp}, we describe our equation set and control parameters. In Section \ref{sec:regimes}, we describe the three solution regimes (weak-, medium-, and strong-field) that our dynamos achieve. In Section \ref{sec:tach}, we present the degree of tachocline confinement, as well as the associated torque balance, for our simulations. In Section \ref{sec:cycle}, we describe the two distinct types of magnetic cycle exemplified by the weak- and strong-field regimes. In Section \ref{sec:skindepth}, we show that for all cases, the poloidal magnetic field strength in the RZ is consistent with diffusive imprinting of the CZ's poloidal field according to the electromagnetic skin effect. In Section \ref{sec:cyclena}, we highlight the distinctions between axisymmetric and non-axisymmetric polarity reversals. Finally, in Section \ref{sec:concl}, we discuss our results in the context of the solar tachocline confinement problem.
 
\section{Numerical Scheme \& Simulation Parameters}\label{sec:exp}
We evolve the 3D magnetohydrodynamic (MHD) equations in spherical shells using the open-source {\rayleigh} code \citep{Featherstone2016a,Matsui2016,Featherstone2021}. We make use of both spherical coordinates [$r$ (radius), $\theta$ (colatitude), and $\phi$ (azimuth angle)] and cylindrical coordinates [$\lambda=r\sin\theta$ (cylindrical radius), $\phi$ (azimuth angle), and $z=r\cos\theta$ (axial coordinate)]. The symbol $\e$ denotes a unit vector. The equations are solved in a frame rotating with the constant angular velocity $\bm{\Omega}_0=\Omega_0\ez$. The Coriolis force is kept but the oblateness and centrifugal force are ignored. Each shell extends from an inner radius $r\inn$ to an outer radius $r\out$. We divide the shell into two layers of equal depth, separated at $r_0\definealt(r\inn+r\out)/2$. The top half ($r_0$ to $r\out$; the CZ) is nominally convectively unstable and the bottom half ($r\inn$ to $r_0$; the RZ) convectively stable. 

{\rayleigh} solves the anelastic MHD equations, which allow significant density contrast across the shell, but disallow sound waves (e.g., \citealt{Ogura1962,Gough1969,Gilman1981,Clune1999}). The anelastic approximation consists of assuming a solenoidal mass flux [see Equation \eqref{eq:contdim}] and thermodynamic perturbations that are small relative to a well-chosen ``background" or ``reference" state. In \rayleigh, the background state is always spherically symmetric and time-independent (e.g., \citealt{Featherstone2016a}). We choose a background entropy gradient $\dsdrline$ that changes from stable to unstable near $r=r_0$ over the transition width $\delta$ and a gravitational acceleration $\gref=G\msun/r^2$ (where $G=\sn{6.67}{-8}\ \rm{cm^3\ g^{-1}\ s^{-2}}$ is the universal gravitational constant and $\msun=\sn{1.99}{33}\ \rm g$ the solar mass). If we further assume a hydrostatic, ideal gas [with constant specific heats $\cv$ (at constant volume) and $\cp$ (at constant pressure)], the choices for $\dsdrline$ and $\gref$ determine the background density $\rhoref$, temperature $\tmpref$, and squared buoyancy frequency $\nsqref\definealt(\gref/\cp)\dsdrline$ (we use $\nsqref$ in favor of $\dsdrline$ in the equations). 

We choose all diffusivities (kinematic viscosity $\nuref$, thermal diffusivity $\kapparef$, and magnetic diffusivity $\etaref$) to increase with height like $1/\rhoref^{1/2}$. We choose an internal heating function $\qref$ (representing radiative heating from below) that deposits thermal energy preferentially in roughly the bottom third of the CZ and drives convection. In the RZ, we set $\qref=0$, tapered from its profile in the CZ over a width $\delta_{\rm{heat}}$. We fully describe our reference state in Appendix \ref{ap:ref} and its analogy to the Sun in Appendix \ref{ap:analog}. 

The dimensional equations of motion are
\begin{align}
	\nabla\cdot(\overline{\rho}\vecu) &=  0,
	\label{eq:contdim}
\end{align}
\begin{align}
	\nabla\cdot\bm{B} &=  0,
	\label{eq:divb0dim}
\end{align}
\begin{subequations}
	\begin{align}
		\overline{\rho}\left(\frac{D\vecu}{Dt} \right) &= -2\rhoref\bm{\Omega}_0\times\vecu-\overline{\rho}\nabla \left(\frac{\prspert}{\overline{\rho}}\right) +\frac{\overline{\rho}\,\gref \spert}{\cp}\er \nonumber\\
		&\ \ \ + \nabla\cdot \bm{D}	+ \frac{1}{\mu}(\curl\bm{B})\times\bm{B}\nonumber\\\label{eq:momdim}\\
		\where D_{ij} &\definealt 2\rhoref\,\nuref \left[e_{ij} - \frac{1}{3}(\Div\vecu) \delta_{ij} \right]\label{eq:dijdim}\\
		\andd e_{ij} &\definealt \frac{1}{2}\left(\pderiv{u_i}{x_j} + \pderiv{u_j}{x_i} \right)\label{eq:eijdim},
	\end{align}
\end{subequations}
\begin{align}
	\overline{\rho}\overline{T}\left(\frac{D\spert}{Dt} \right) =\ &Q  - \rhoref\tmpref\dsdr u_r + \nabla\cdot\left(\kapparef\,\overline{\rho}\overline{T}\nabla \spert\right]\nonumber \\
	&+ D_{ij}e_{ij} +  \frac{\eta}{4\pi}|\curl\bm{B}|^2,
	\label{eq:endim}
\end{align}
and
\begin{align}
	\pderiv{\bm{B}}{t}  =\ &\curl (\vecu\times\bm{B} - \eta\curl\bm{B}). 
	\label{eq:inddim}
\end{align}
Here, $D/Dt\definealt\pderivline{}{t}+\ugrad$ is the material derivative and $\mu$ the vacuum permeability ($\mu=4\pi$ in Gaussian units). 

 {\rayleigh} was originally run by solving these dimensional equations. In this work, however, we discuss only the equivalent non-dimensional simulations. Length is scaled by the CZ (or RZ) thickness $H\definealt(r\out-r\inn)/2$ and time by the rotational time-scale $\Omega_0^{-1}$. The velocity $\vecu$ is scaled by $[\vecu]\definealt\Omega_0H$ and the vorticity $\vecom\definealt\curl\vecu$ by $\Omega_0$ (we use square brackets to denote the unit of each fluid variable). Each background-state profile is scaled by its volume-average over the CZ (denoted by a tilde, e.g., $\tilde{\rho}$), except for $\nsqref$, which is scaled by its volume-average over the RZ (denoted by $\langle \overline{N^2}\rangle\rz$), and $\qref$, which is scaled as described below. The pressure perturbation $\prspert$ is scaled by $[P]\definealt\tilde{\rho}(\Omega_0H)^2$ and the magnetic field $\vecb$ by $[\vecb]\definealt \sqrt{\mu\tilde{\rho}}(\Omega_0H)$.

As noted by \citet{Christensen2006}, the chosen non-dimensionalization omits the diffusivities from the scales for time and the magnetic field. This is helpful in extending scaling relationships to stellar regimes, where diffusive effects are not believed to play a large role (although we note at the outset that such scaling relationships are likely not present in this work, where diffusive effects \textit{do} play a large role). An added benefit of this non-dimensionalization is that $\vecu$ and $\vecb$ appear with order-unity coefficients in the momentum equation, so their relative importance (to both the force balance and the partition of kinetic and magnetic energy) can be inferred directly from their non-dimensional values. 

The internal heating $\qref$, coupled with the thermal boundary conditions described below, drives convection by establishing sharp entropy gradients in a thermal boundary layer near the top of the CZ (e.g., \citealt{Featherstone2016a, Matilsky2020b}). This convection (and conduction, especially in the boundary layer), must carry a ``non-radiative" energy flux $\fluxnr\definealt (1/r^2)\int_{r_0}^r\qref(x)x^2dx$ in the statistically steady state. The entropy perturbation $\spert$ is thus scaled by its estimated difference across the conductive boundary layer ($[S]=\Delta S\definealt\fluxnrtilde H/\tilde{\rho}\tilde{T}\tilde{\kappa}$) and $\qref$ by $\fluxnrtilde/H$. 

With these scaling choices, the non-dimensional equations of motion are
	\begin{align}\label{eq:mom}
		\rhoref\left(\matderiv{\vecu}\right) &= -2\rhoref\ez\times\vecu-\rhoref\nabla\left(\frac{\prspert}{\rhoref} \right) +\rafmod \rhoref\, \gref \spert\er \nonumber\\
		&\ \ \ +\ek \Div\bm{D}+(\curl\vecb)\times\vecb
	\end{align}
\begin{align}\label{eq:heat}
	\rhoref\tmpref \matderiv{\spert} = &\frac{\ek}{\pr} \qref - \frac{{\rm{Bu}}}{\rafmod} \rhoref\tmpref \frac{\nsqref}{\gref} \urad+ \frac{\ek}{\pr} \Div(\rhoref \tmpref \kapparef \nabla \spert)  \nonumber\\
	&+ \frac{\di\ek}{\rafmod} D_{ij}e_{ij} + \frac{\di\ek}{\prm\rafmod} \etaref|\curl\vecb|^2,
\end{align}
\begin{align}\label{eq:ind}
	\andd \pderiv{\vecb}{t} = \curl(\vecu\times\vecb) - \frac{\ek}{\prm} \curl(\etaref\curl\vecb).
\end{align}	
 \newtext{Here, Equations \eqref{eq:contdim} and \eqref{eq:divb0dim} still apply and are unchanged, and $D_{ij}$ and $e_{ij}$ are defined exactly as in Equations \eqref{eq:dijdim} and \eqref{eq:eijdim}}, respectively. All field variables ($\vecu$, $\vecb$, $\spert$, and $\prspert$), spatial quantities ($r$, $t$, $\lambda$, $z$, and $\nabla$), and background-state profiles now denote their non-dimensional values. The non-dimensional input numbers (definitions and values) are given in Table \ref{tab:inputnondim}. 
 
 The reference-state control parameters are the ratio of specific heats $\gamma$, the CZ-to-RZ aspect ratio $\alpha$, the CZ aspect ratio $\beta$, the number of scale heights across the CZ $N_\rho$, and the transition widths $\delta$ and $\delta_{\rm{heat}}$. This reference state (except for the diffusivity profiles) is reasonably solar-like and describes the upper 2.1 density scale-heights of the solar RZ and the lower 3 density scale-heights of the solar CZ (see Appendix \ref{ap:ref}). In units of $H$, the non-dimensional solar radius is $R_\odot=4.39$ (see Table \ref{tab:solaranalog}). We plot radial profiles as functions of $r/R_\odot$, to more easily compare to prior work.
 
The fluid control parameters are the Prandtl number $\pr$, the magnetic Prandtl number $\prm$, the modified Rayleigh number $\rafmod$, the Ekman number $\ek$, and the buoyancy number $\rm{Bu}$. The dissipation number $\di\definealt \tilde{g}H/(\cp\tilde{T})=1.72$ for the cases here and is not a control parameter in our convention, being a function of $\gamma$, $\beta$, and $N_\rho$, which we deem reference state control parameters (see Appendix \ref{ap:ref} and \citealt{Korre2021}). Some additional parameters (that can be derived from the input parameters given in Table \ref{tab:inputnondim}) are given in Table \ref{tab:inputnondimderiv}.

Equations \eqref{eq:contdim}--\eqref{eq:inddim} are discretized in space. For all simulations, we use three sets of stacked Chebyshev collocation points in $r$ ($N_r/3=64$ points in each domain), $N_\theta=384$ Legendre collocation points in $\theta$, and  $N_\phi=2N_\theta=768$ uniformly spaced collocation points in $\phi$. The Chebyshev points cluster near each domain's boundaries. We require increased resolution in the overshoot layer (i.e., in the vicinity of $r_0/\rsun=0.719$), and so we set the radial domain boundaries to lie at $r/\rsun =\{0.491, 0.669, 0.719, 0.947\}$ (or equivalently, $r-r_0=\{-1.000, -0.219, 0.000, 1.000\}$). Nonlinear terms and the Coriolis force are evaluated in physical space (i.e., on the discretized spatial grid), while the remaining linear terms are evaluated in spectral space, using Chebyshev polynomials in each $r$ sub-domain and spherical harmonics in $\theta$ and $\phi$. The variables in physical space are de-aliased using the 2/3 rule: the maximum Chebyshev degree (in each $r$ sub-domain) is $n_{\rm{max}}=42$ and the maximum spherical harmonic degree is $\lmax=255$. For more details, see \citet{Glatzmaier1984} and \citet{Clune1999}, who pioneered {\rayleigh}'s pseudo-spectral algorithm.

\begin{table}
	\caption{Non-dimensional control parameters for our simulations. We list reference-state parameters first, then the fluid control parameters.}
	\label{tab:inputnondim}
	\centering
	\begin{tabular}{l  l l}
		\hline
		Parameter & Definition & Value\\
		\hline
		$\gamma$ & $\cp/\cv$  & $5/3$\\
		$\alpha$ & $(r\out-r_0)/(r_0-r\inn)$  &  1\\
		$\beta$ & $ r_0/r\out$ & 0.759\\
		$N_\rho$ & $\ln[\rhoref(r_0)/\rhoref(r\out)]$ & 3.00\\
		$\delta$ & stability transition width & 0.219\\
		$\delta_{\rm{heat}}$ & heating transition width & 0.132\\
		\hline
		$\pr$ & $\tilde{\nu}/\tilde{\kappa}$ & 1\\
		$\prm$ & $\tilde{\nu}/\tilde{\eta}$ & 1 to 8\\
		$\rafmod$ & $\fluxnrtilde \tilde{g}/(\cp\tilde{\rho}\tilde{T}\tilde{\kappa}\Omega^2)$ &  0.638\\
		$\ek$ & $ \tilde{\nu}/(\Omega_0H^2)$ & $\sn{1.07}{-3}$ \\
		$\rm Bu$ & $\av{N^2}\rz/\Omega_0^2$ & $\sn{2.54}{4}$\\[1ex]
		\hline
	\end{tabular}
\end{table}

Each magnetic simulation differs only in the choice of $\prm$, which ranges from 1 to 8. We also consider a purely hydrodynamic simulation (referred to as ``Case H"), which has all the parameters listed in Table \ref{tab:inputnondim}, but no magnetic field. We refer to each magnetic case by its value of $\prm$ rounded to two decimal places: e.g., ``Case 1.08" means $\prm=1.076$. Cases H and 4.00 were analyzed in \citetalias{Matilsky2022}. All chosen values of $\prm$ are listed in Table \ref{tab:outputnondim}.

At both boundaries, we use stress-free and impenetrable conditions on $\vecu$, potential-field-matching conditions on $\vecb$, and fixed-entropy-gradient conditions on $\spert$. Specifically, we set $\pderivline{\spert}{r}$ to zero at the bottom boundary (thus allowing no conductive flux in or out) and set it to a latitudinally independent negative value at the top boundary, such that the energy conducted out the top is equal to the energy injected by $\qref$ (e.g., \citealt{Matilsky2020b}). The convection is initialized by introducing weak noise in $\spert$ (amplitude $\sim$$10^{-3}$), randomly distributed in space throughout the entire shell. For the magnetic cases, we further introduce weak noise in $\vecb$ (amplitude $\sim$$10^{-6}$), randomly distributed in space throughout the CZ only. The other field variables ($\vecu$ and $\prspert$) are initialized to zero in all space. 

We use several types of averages in this work. Let $\psi=\psi(r,\theta,\phi,t)$ denote a scalar quantity (or a single component of a vector quantity) dependent on position and time. Then $\avphi{\psi}$, $\avsph{\psi}$, $\av{\psi}\cz$, $\av{\psi}\rz$, and $\av{\psi}\full$ denote instantaneous averages of $\psi$ over longitude, spherical surfaces, the CZ (volume-average from $r_0$ to $r\out$), the RZ (volume-average from $r\inn$ to $r_0$), and the full shell (volume-average from $r\inn$ to $r\out$), respectively.  An additional temporal average (over the ``equilibrated state"; see the following section) is denoted by appending a ``$t$" to the subscript in the average: e.g., $\avphit{\psi}$. Subtracting the instantaneous longitudinal average is denoted by a prime: $\psi^\prime\definealt \psi-\avphi{\psi}$. We also colloquially refer to $\avphi{\psi}$ and $\psi^\prime$ as the ``mean and fluctuating" components of $\psi$, respectively.
\begin{figure*}
	\centering
	\includegraphics[width=7.25in]{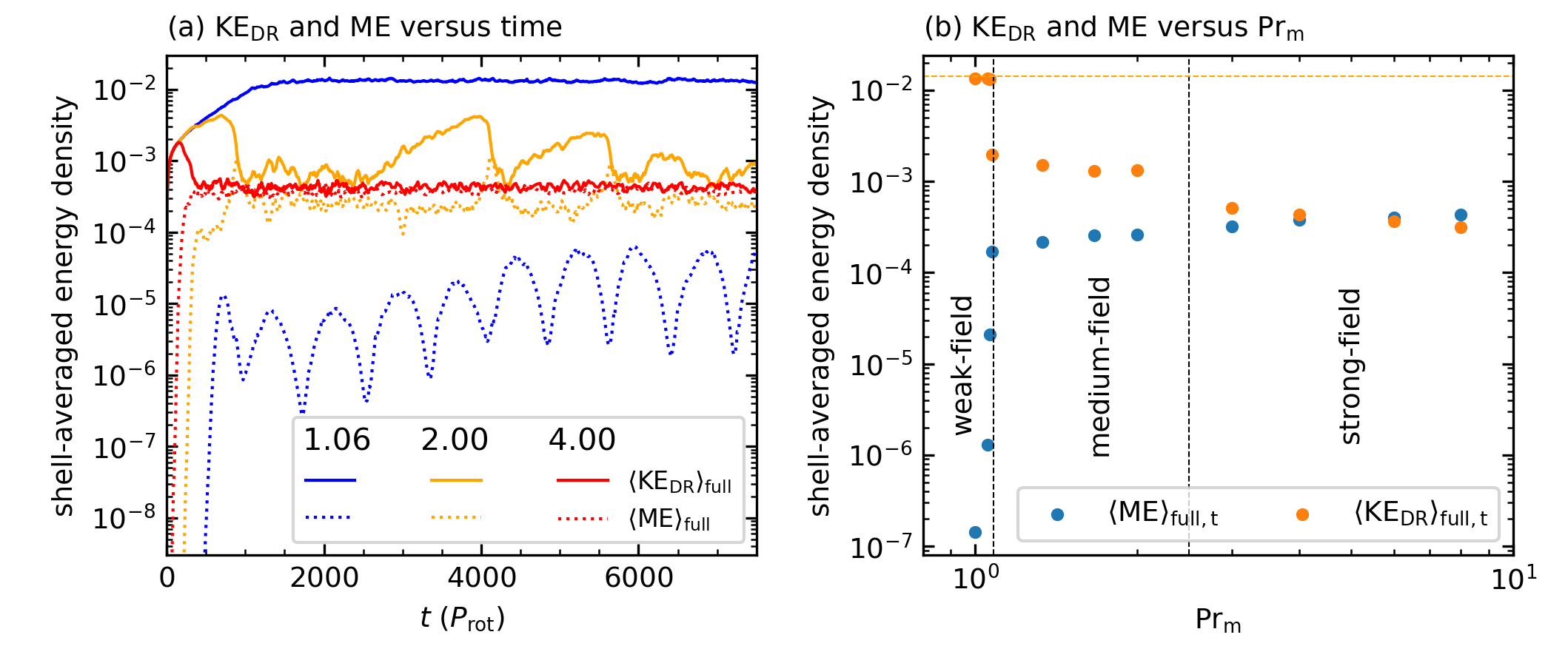}
	\caption{Averaged $\rm KE_{DR}$ and $\rm ME$ as functions of time and $\prm$. (a)  $\av{\rm{KE_{DR}}}\full$ (solid curves) and $\av{\rm{ME}}\full$ (dashed curves) with respect to time for three values of $\prm$ (indicated by three different line colors and legend headings). (b)  $\av{\rm{KE_{DR}}}_{\rm full,t}$ and $\av{\rm{ME}}_{\rm full,t}$ with respect to $\prm$ for all magnetic simulations. Vertical black lines denote tentative regime boundaries and the horizontal orange line marks $\av{\rm{KE_{DR}}}_{\rm full,t}$ for Case H.}
	\label{fig:energy}
\end{figure*}

\section{Dynamo Regimes}\label{sec:regimes}
All the magnetic cases presented here yield sustained large-scale dynamos. As convection and dynamo action become significant, the field variables grow from their initially small values to amplitudes of order unity. We quantify this growth in terms of the kinetic energy density of the differential rotation, $\rm KE_{DR}$, and the magnetic energy density, $\rm ME$:
\begin{align}\label{eq:drke_and_me}
{\rm KE_{DR}} \definealt \frac{1}{2}\rhoref\avphi{u_\phi}^2 \five\andd  {\rm ME}\definealt \frac{1}{2}\avphi{\vecb^2}.
\end{align} 

\newtext{In Equation \eqref{eq:drke_and_me}, the energy densities are functions of $r$, $\theta$, and $t$, which we average further in the subsequent analysis.} In Figure \ref{fig:energy}, we show the growth and long-term behavior of the \newtext{full-shell-averaged} energy densities for some representative simulations. After a certain time (which we call $t=\tequil$), the system achieves a ``statistically steady" or ``equilibrated" state, in which the volume-averaged magnitude of each field variable fluctuates about a well-defined temporal mean. We choose $\tequil$ (fairly roughly) by eye from plots like Figure \ref{fig:energy}(a). For example, we choose $\tequil=2000\prot$ for Case 1.06, $\tequil=1000\prot$ for Case 2.00, and $\tequil=600\prot$ for Case 4.00 (see Table \ref{tab:outputnondim} for all values of $\tequil$).

Figure \ref{fig:energy}(a) suggests three basic dynamo regimes. The low-$\prm$ solution (Case 1.06; blue curves) lies in a ``weak-field regime", characterized by $\av{\rm ME}\full$ always being orders of magnitude weaker than $\av{\rm KE_{DR}}\full$. There is a regular magnetic energy cycle, with a period of roughly $750\ \prot$. The high-$\prm$ solution (Case 4.00; red curves) lies in a ``strong-field regime", characterized by $\av{\rm ME}\full$ about in equipartition with $\av{\rm KE_{DR}}\full$, while $\av{\rm KE_{DR}}\full$ itself is much weaker than in the weak-field case. Finally, the intermediate-$\prm$ solution (Case 2.00; orange curves) lies in a ``medium-field regime." Case 2.00 has properties similar to those of a strong-field dynamo some of the time, but occasionally $\av{\rm ME}\full$ falls below its strong-field value and then $\av{\rm KE_{DR}}\full$ steadily increases above its strong-field value (representing an increase in differential rotation or partial disappearance of the tachocline). After $\av{\rm KE_{DR}}\full$ reaches a critical level, $\av{\rm ME}\full$ grows rapidly, lowering $\av{\rm KE_{DR}}\full$ back to its lower, strong-field value. There is no clear cycling behavior, obvious physical trigger, or general predictability for the medium-field cases' temporary epochs of strong differential rotation.  

Figure \ref{fig:energy}(b) shows the equilibrated levels of the kinetic energy in the differential rotation, $\av{\rm KE_{DR}}_{{\rm full},t}$, and the magnetic energy, $\av{\rm ME}_{{\rm full},t}$, for all simulations. The weak-field regime (for which the differential rotation has the same magnitude as in Case H) sits in the narrow range of roughly $1.00\lesssim\prm\lesssim1.06$. The medium-field regime (for which the differential rotation is substantially weakened compared to the weak-field cases but intermittently becomes stronger) occupies roughly $1.08\lesssim\prm\lesssim2.5$. The strong-field regime (lowest differential rotation and highest magnetic energy) occupies $\prm\gtrsim2.5$. Note that these identified regimes and their boundaries are only suggestive, given our limited resolution in $\prm$-space. 

\begin{figure*}
	\centering
	\includegraphics[width=7.25in]{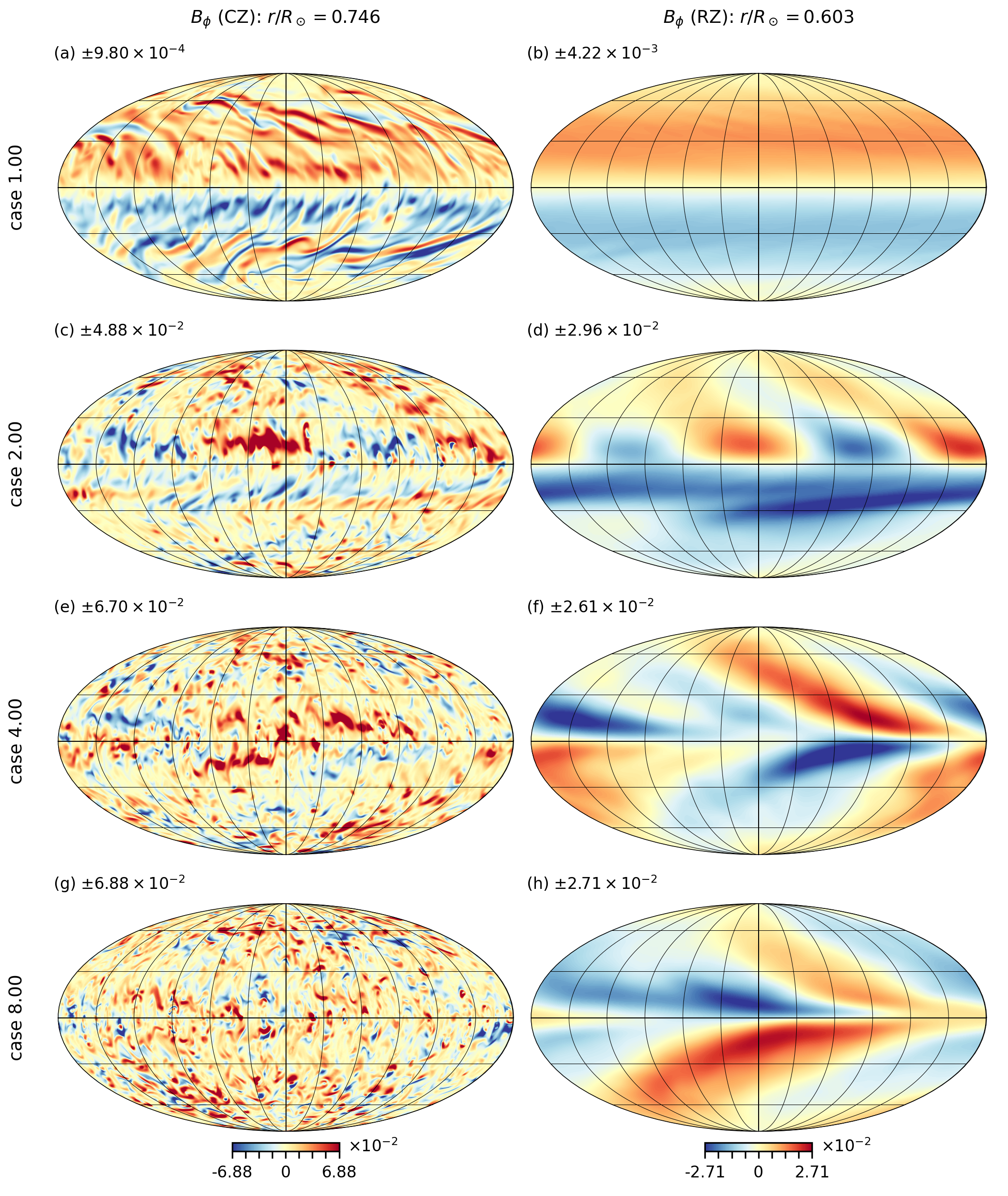}
	\caption{Mollweide projections of the toroidal magnetic field $B_\phi$ on spherical surfaces for four chosen values of $\prm$ at time $t=3500\prot$. Each $\prm$ corresponds to a different row (pair) of Mollweides, and $\prm$ increases downward. The spherical surfaces are at two radii, one near the base of the CZ (left-hand column) and one in the middle of the RZ (right-hand column). The colorbar (shown for the bottom row only) is the same for all figures and we give its saturation values next to the alphabetical labels.}
	\label{fig:sslice}
\end{figure*}

The weak-field dynamos tend to be more axisymmetric (magnetism dominated by azimuthal wavenumber $m=0$) than the strong-field dynamos. Figure \ref{fig:sslice} shows the toroidal magnetic field projected on spherical surfaces for four solutions at different $\prm$ and therefore in the different regimes. For Case 1.00 (the weak-field regime), there is a strong $m=0$ component, both in the CZ and even more so in the RZ. For higher values of $\prm$ (the medium- and strong-field regimes), the field in the CZ becomes increasingly dominated by small scales, but retains a large-scale ($m=0,1,2$) envelope. For all cases, the RZ appears to act as a low-pass filter for the spatial scales of the field, letting only the low $m$'s survive. This is especially apparent for Case 8.00 [Figures \ref{fig:sslice}(g,h)]. 

As a whole, the dominant $m=0$ field structures at low $\prm$ resemble what has previously been called magnetic ``wreaths"---toroidal bands of strong magnetism looping the full sphere in a given hemisphere (e.g., \citealt{Brown2010,Passos2014,Bice2020}). Such wreaths are often invoked in connection to the magnetic butterfly diagram, as interior reservoirs of toroidal field from which smaller loops can potentially break off and buoyantly rise to form sunspot pairs (e.g., \citealt{DSilva1993,Stenflo2012,Nelson2013a,Li2018,Bice2023}). 

For higher $\prm$, the strong $m=1,2$ components of $\vecb$ resemble the ``partial wreaths" discussed at some length by \citet{Matilsky2020a,Matilsky2020c}. The cases from that work contained a dominant $m=1$ field structure that appeared to be two opposite-polarity full wreaths tilted into each other, or possibly linked. On a spherical slice, the tilted full wreaths showed up as two opposite-polarity ``partial wreaths", extending in longitude by about $180^\circ$, with central longitudes on opposite sides of the sphere. We should note that ``partial wreath" is really a placeholder for lack of a better term. The 3D field-line tracings of these non-axisymmetric structures tend to be quite difficult to interpret and it remains unclear exactly what topology (linked wreaths, tilted wreaths, or even open field-lines) is leading to the two-dimensional projections shown in Figure \ref{fig:sslice}. 

To quantify the non-axisymmetry in our dynamos more precisely, we partition the magnetic energy according to $m$-value. We define the $m$-component of $\vecb$ through
\begin{subequations}\label{eq:bm}
	\begin{align}
		\vecb _m &\equiv \avphi{\vecb e^{-im\phi}}\label{eq:mcomp1}\\
		\orr \vecb &\equiv \sum_m \vecb_m e^{im\phi}.\label{eq:mcomp2}
	\end{align}
\end{subequations}
Note that since $\avphi{\cdots}$ is computed by averaging over the uniformly spaced $\phi$-grid, Equation \eqref{eq:bm} represents the forward and inverse discrete Fourier transforms (DFTs) in $\phi$. For the chosen normalization, \newtext{Parseval's} theorem takes the form
\begin{align}\label{eq:parceval}
	\avphi{\vecb^2} = \sum_m |\vecb_m|^2.
\end{align}
Note that by definition, $\vecb_0=\avphi{\vecb}$. Different components of the magnetic energy can thus be attributed to different $m$-components of $\vecb$. 

\begin{table*}
	\caption{Basic simulation properties ($\prm$, regime, run time $\tmax$, and equilibration time $\tequil$) and the partition of kinetic and magnetic energy for each simulation's CZ and RZ. Here, we define the kinetic energy of the convection (fluctuating flows) as ${\rm KE_c}\definealt (1/2)\rhoref \avphi{(\vecu^\prime)^2}$. \newtext{Recall that the magnetic energy in the fluctuating fields is defined through ${\rm ME_{\geq3}}\definealt (1/2)\sum_{|m|\geq3}|\vecb_m|^2$.} The diffusion time $P_{\rm{diff}}$ refers to the viscous (or equivalently, thermal) diffusion time $\pnu=\pkappa$ for Case H and to the magnetic diffusion time $\peta$ for the magnetic cases (see Table \ref{tab:inputnondimderiv}).  The letters ``W," ``M," and ``S," denote the weak-, medium-, and strong-field regimes, respectively.  For the energy ratios, a volumetric (over the CZ or RZ) and temporal mean is implied for the numerator and denominator separately. For example, ``$2|\vecb_1|^2/\vecb^2$" in the ``CZ block" of the table should be read as $2\av{|\vecb_1|^2}_{{\rm CZ},t}/\av{\vecb^2}_{{\rm CZ}, t}$. The factors of 2 account for the symmetry $|\vecb_{-m}|^2 = |\vecb_m|^2$ (since $\vecb$ is real).} 
	\label{tab:outputnondim}
	\centering
	\begin{tabular}{*{13}{l}  }
		\hline
		Case & H & 1.00 & 1.05 & 1.06  &  1.08     & 1.33 &  1.67 & 2.00  & 3.00 & 4.00 & 6.00 & 8.00 \\
		\hline
		
$\prm$   &   -   &   1.000   &   1.054   &   1.065   &   1.076   &   1.333   &   1.667   &   2.000   &   3.000   &   4.000   &   6.000   &   8.000  \\
regime   &   -   &   W   &   W   &   W   &   M   &   M   &   M   &   M   &   S   &   S   &   S   &   S  \\
$t_{\rm{eq}}/P_{\rm{rot}}$   &   1000   &   1500   &   1500   &   2000   &   2300   &   1500   &   1200   &   1000   &   1700   &   600   &   500   &   500  \\
$t_{\rm{max}}/P_{\rm{rot}}$   &   9930   &   9740   &   6670   &   7770   &   9400   &   7480   &   8170   &   9200   &   7730   &   16000   &   5450   &   5750  \\
$t_{\rm{max}}/P_{\rm{diff}}$   &   12.7   &   12.5   &   8.13   &   9.36   &   11.2   &   7.20   &   6.29   &   5.90   &   3.31   &   5.15   &   1.17   &   0.92  \\
\hline
\multicolumn{13}{c}{CZ energy density parameters}\\
$\rm KE_{DR}$   &   0.011   &   0.011   &   0.011   &   0.011   &   2.05e-3   &   1.83e-3   &   1.24e-3   &   1.25e-3   &   4.70e-4   &   3.75e-4   &   2.82e-4   &   2.19e-4  \\
$\rm KE_c$   &   1.71e-3   &   1.71e-3   &   1.69e-3   &   1.69e-3   &   9.71e-4   &   9.52e-4   &   9.35e-4   &   9.31e-4   &   9.08e-4   &   8.90e-4   &   8.74e-4   &   8.58e-4  \\
$\rm ME$   &   -   &   1.99e-8   &   1.82e-7   &   3.16e-6   &   1.56e-4   &   2.08e-4   &   2.72e-4   &   2.94e-4   &   4.21e-4   &   5.08e-4   &   5.52e-4   &   5.98e-4  \\
$\avphi{\vecb}^2/\vecb^2$   &   -   &   0.389   &   0.371   &   0.334   &   0.088   &   0.108   &   0.077   &   0.066   &   0.028   &   0.021   &   0.017   &   0.013  \\
$2|\vecb_1|^2/\vecb^2$   &   -   &   0.042   &   0.042   &   0.043   &   0.248   &   0.156   &   0.144   &   0.117   &   0.121   &   0.106   &   0.064   &   0.044  \\
$2|\vecb_2|^2/\vecb^2$   &   -   &   0.040   &   0.042   &   0.067   &   0.165   &   0.179   &   0.162   &   0.149   &   0.091   &   0.063   &   0.046   &   0.041  \\
$\rm ME_{\geq3}/ME$   &   -   &   0.529   &   0.545   &   0.556   &   0.499   &   0.556   &   0.617   &   0.668   &   0.760   &   0.809   &   0.874   &   0.902  \\
\hline
\multicolumn{13}{c}{RZ energy density parameters}\\
$\rm KE_{DR}$   &   0.016   &   0.016   &   0.016   &   0.016   &   1.20e-3   &   1.30e-3   &   8.97e-4   &   9.40e-4   &   5.33e-5   &   3.62e-5   &   3.17e-5   &   2.38e-5  \\
$\rm KE_c$   &   4.14e-4   &   4.14e-4   &   4.00e-4   &   4.09e-4   &   5.37e-5   &   4.03e-5   &   3.12e-5   &   2.98e-5   &   2.28e-5   &   2.18e-5   &   2.03e-5   &   1.90e-5  \\
$\rm ME$   &   -   &   3.71e-7   &   3.41e-6   &   5.47e-5   &   1.97e-4   &   2.26e-4   &   2.28e-4   &   2.09e-4   &   1.36e-4   &   1.38e-4   &   1.34e-4   &   1.28e-4  \\
$\avphi{\vecb}^2/\vecb^2$   &   -   &   0.957   &   0.955   &   0.957   &   0.527   &   0.658   &   0.603   &   0.587   &   0.117   &   0.076   &   0.127   &   0.096  \\
$2|\vecb_1|^2/\vecb^2$   &   -   &   5.01e-3   &   5.02e-3   &   5.10e-3   &   0.320   &   0.151   &   0.161   &   0.149   &   0.399   &   0.446   &   0.397   &   0.359  \\
$2|\vecb_2|^2/\vecb^2$   &   -   &   4.21e-3   &   4.33e-3   &   4.53e-3   &   0.098   &   0.132   &   0.162   &   0.172   &   0.273   &   0.242   &   0.210   &   0.249  \\
$\rm ME_{\geq3}/ME$   &   -   &   0.034   &   0.035   &   0.033   &   0.054   &   0.059   &   0.074   &   0.092   &   0.210   &   0.236   &   0.266   &   0.296  \\

		\hline
	\end{tabular}
	
\end{table*}

Table \ref{tab:outputnondim} shows some basic simulation properties ($\prm$, regime, total run time $\tmax$, and equilibration time $\tequil$), as well as the partitions of kinetic and magnetic energy for each simulation.\footnote{For completeness and reproducibility, we also give the Reynolds, Rossby, and magnetic Reynolds numbers for each case in Appendix \ref{ap:outputnd}.} We define the magnetic energy in the ``small-scale fields" by ${\rm ME_{\geq3}}\definealt (1/2)\sum_{|m|\geq3}|\vecb_m|^2$. With increasing $\prm$, the fraction of magnetic energy in the small-scale fields ($\rm ME_{\geq3}/ME$) increases, as might be expected from the more prominent small-scale structures seen in Figure \ref{fig:sslice} at higher $\prm$. The deficit, i.e., the fraction of energy in the large-scale fields ($m=0,1,2$) decreases, but the partition between each $m$-component is complex. 

For the weak-field cases, in both the CZ and RZ, the power in the axisymmetric field dominates over the power in the $m=1,2$ components. But for the medium- and strong-field cases, there seems to be no general rule for whether $m=0,1,$ or $2$ dominates. One robust feature is that in all regimes, the magnetic energy in the RZ is stored primarily in the large-scale fields. Even for the strongest-field Case 8.00, the small-scale ($|m|\geq3$) field components account for only $\sim$30\% of the magnetic energy in the RZ. These results strengthen the earlier idea that the RZ acts as a low-pass filter, letting in only the lowest-$m$ components of the field. Such behavior is expected if the field evolution in the RZ is primarily governed by diffusion, an idea we return to in Section \ref{sec:skindepth}, where we discuss the skin-effect behavior of the poloidal field.  
\begin{figure*}
	\centering
	\includegraphics[width=7.25in]{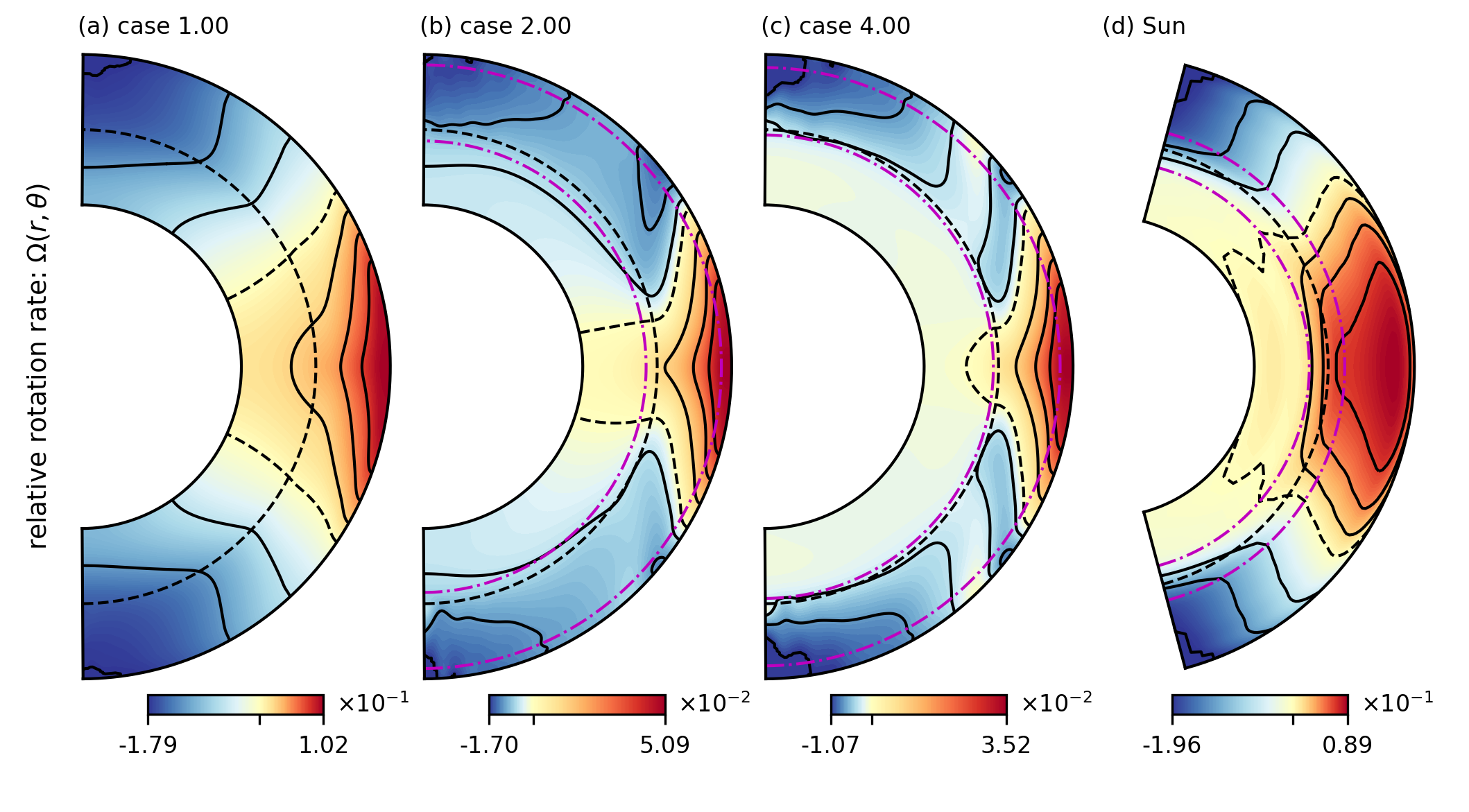}
	\caption{Relative rotation rate $\Omega$ for (a)--(c) simulations in each of the three dynamo regimes and (d) the Sun, plotted in color in the meridional plane. Each colormap is bi-linear: positive values (red tones) are normalized separately from negative values (blue tones). The minimum and maximum saturation ticks are labeled on the colorbar, while the zero tick is unlabeled. Overplotted, there are three equally spaced positive and negative solid contours. The zero contour is dashed. The dashed black curves shows the location of $r_0$ and for (b)--(d), the dash-dotted magenta curves show the boundaries of the tachocline, $r_t\pm \Gamma/2$. The solar rotation rate is from a helioseismic inversion of GONG data averaged from 1995 to 2009 \citep{Howe2005,Howe2023}. To arrive at the non-dimensional, relative rotation rate $\Omega$  for the Sun, we define $\Omega_\odot\definealt\sn{2.70}{-6}\ {\rm rad\ s^{-1}}$ (or $\Omega_\odot/2\pi=430$ nHz), which is roughly the solid-body rotation rate of the solar RZ. We then subtract $\Omega_\odot$ from the inverted rotation rate (which is given dimensionally, in the non-rotating frame) and divide by $\Omega_\odot$.}
	\label{fig:dr}
\end{figure*}
\begin{figure*}
	\centering
	\includegraphics[width=7.25in]{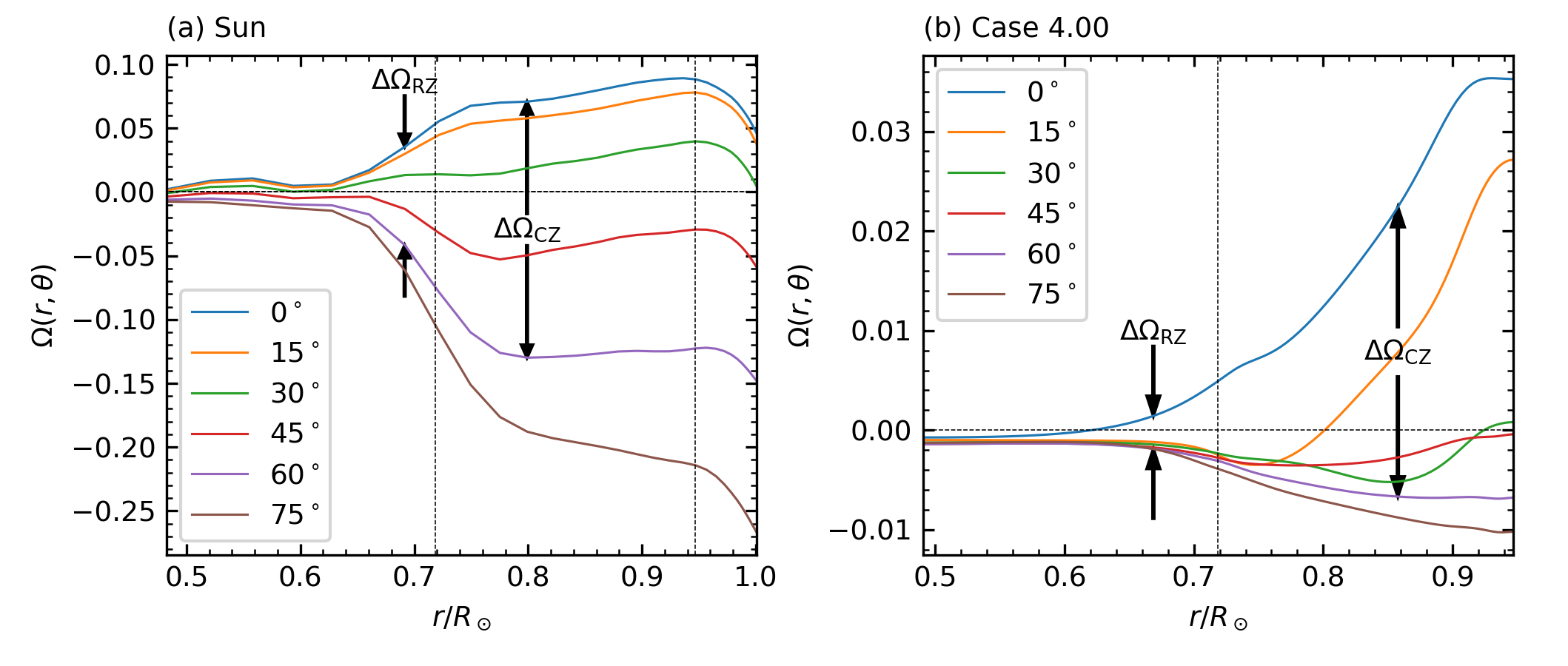}
	\caption{Relative rotation rate $\Omega$, plotted along radial lines for (a) the Sun and (b) the strong-field Case 4.00. Six radial cuts of $\Omega$ are plotted, equally spaced in latitude by $15^\circ$ between $0^\circ$ and $75^\circ$. In panel (a), the $x$-axis is extended slightly, since the simulations only extend to $r\out=0.947\rsun$. In both panels, the vertical arrows represent the values of $\Delta\Omega\rz$ and $\Delta\Omega\cz$. In this figure (and in all radial plots in subsequent figures), the thin vertical lines denote the locations of $r_0$ and $r\out$. }
	\label{fig:drline}
\end{figure*}

\section{Tachocline Confinement}\label{sec:tach}
\subsection{Tachocline Appearance}\label{sec:tach1}
All dynamos in the medium- and strong-field regimes sustain steady-state tachoclines. To describe them, we define the rotation rate $\Omega$ (as measured in the rotating frame):
\begin{align}\label{eq:rotrate}
	\Omega(r,\theta)\equiv \frac{\avphit{u_\phi}}{\lambda}.
\end{align}
 Figure \ref{fig:dr} shows $\Omega$ for a weak-, medium-, and strong-field case, as well as the Sun. The weak-field case does not have a tachocline and has rotation rate nearly identical to that of Case H, that is, there is strong latitudinal rotation contrast in the CZ ($\Delta\Omega\cz\sim0.2$, similar to the solar value), which imprints throughout the entire RZ (for Case H's rotation profile, see \citetalias{Matilsky2022}). 
 
By contrast, the medium- and strong-field cases all have tachoclines, that is, there is (weak) differential rotation in the CZ, but nearly solid-body rotation in the RZ. The radial transition from differential to solid-body rotation appears to be quite abrupt at most latitudes from the color plots in Figures \ref{fig:dr}(b,c), suggesting thin simulated tachoclines. However, this visual abruptness is at least partly due to our choice of bi-linear colormap for the asymmetric values of $\Omega$ about 0. This colormap deepens the blue tones in the CZ at high latitudes. As we now demonstrate, the simulated tachoclines (after they are fit systematically) are in fact about twice as thick as the solar one. 

We define the radially varying latitudinal differential rotation contrast $\Delta\Omega\ofr$:
\begin{align}\label{eq:drotrate}
	\Delta\Omega(r)\equiv \Omega(r,\pi/2) - \frac{1}{2}[\Omega(r,\pi/6) + \Omega(r,5\pi/6)],
\end{align}
i.e., the difference in rotation rate between the equator and the average rate of 60$^\circ$ latitude north and south. We define the rotation contrasts in the CZ and RZ, and their ratio:
\begin{subequations}\label{eq:dom}
	\begin{align}
		\Delta\Omega\cz&\definealt \av{\Delta\Omega}\cz,\\
		\Delta\Omega\rz&\definealt \av{\Delta\Omega}\rz,\\
		\andd f& \definealt \frac{ \Delta\Omega\rz}{ \Delta\Omega\cz},
	\end{align}
\end{subequations}
respectively. Because the medium- and strong-field RZs often rotate like solid bodies, we define the (volume-averaged) constant rotation rate of the RZ:
\begin{align}\label{eq:omrz}
	\Omega\rz\definealt \av{\Omega}\rz
\end{align}

Figure \ref{fig:drline} shows line plots of the rotation rate along radial lines for the Sun and Case 4.00. Clearly in the solar case, the tachocline is confined to a relatively narrow radial layer, with strong differential rotation in most of the CZ and very little in the RZ, indicating a large $\Delta\Omega\cz$, a small $\Delta\Omega\rz$, and therefore small $f$ for the true solar case. By contrast, in Case 4.00, while $\Delta\Omega_{\rm{RZ}}$ is severely diminished, thus indicating that the RZ rotates nearly as a solid body, $\Delta\Omega_{\rm{CZ}}$ is also significantly diminished. 

A further deviation from the solar case is that our simulated rotation profiles have most of the differential rotation contrast confined to a low-latitude band between about $\pm30^\circ$. The result is relatively strong radial shear distributed far more evenly throughout the CZ in the simulations than in the Sun. Equivalently, each simulated tachocline is not thin, but basically occupies the whole convective layer and is centered near mid-CZ, well above $r_0$.

In order to define the location and width of the simulated tachoclines, we define:
\begin{align}\label{eq:psi}
	\psi(r) &\definealt \frac{\Delta\Omega(r) - {\rm min}(\Delta\Omega)}{{\rm max}(\Delta\Omega) - {\rm min}(\Delta\Omega)} - \frac{1}{2}.
\end{align}
The shape function $\psi(r)$ is normalized to vary between $-1/2$ where $\Delta\Omega$ obtains its minimum value (always in the RZ) and $+1/2$ where $\Delta\Omega$ obtains its maximum value (always in the CZ). We define a given tachocline's centroid $r_t$ and thickness $\Gamma$ as the parameters in the function $(1/2)\tanh[2(r-r_t)/\Gamma]$ which is the best fit to $\psi(r)$.\footnote{This fitting procedure is similar to (though not as involved as) conventional tachocline fitting methods (e.g., \citealt{Charbonneau1999,Basu2003}). We do not feel our simulated rotation profiles are sufficiently solar-like to warrant more complex fitting.}

\begin{figure}
	\centering
	\includegraphics[width=3.6in]{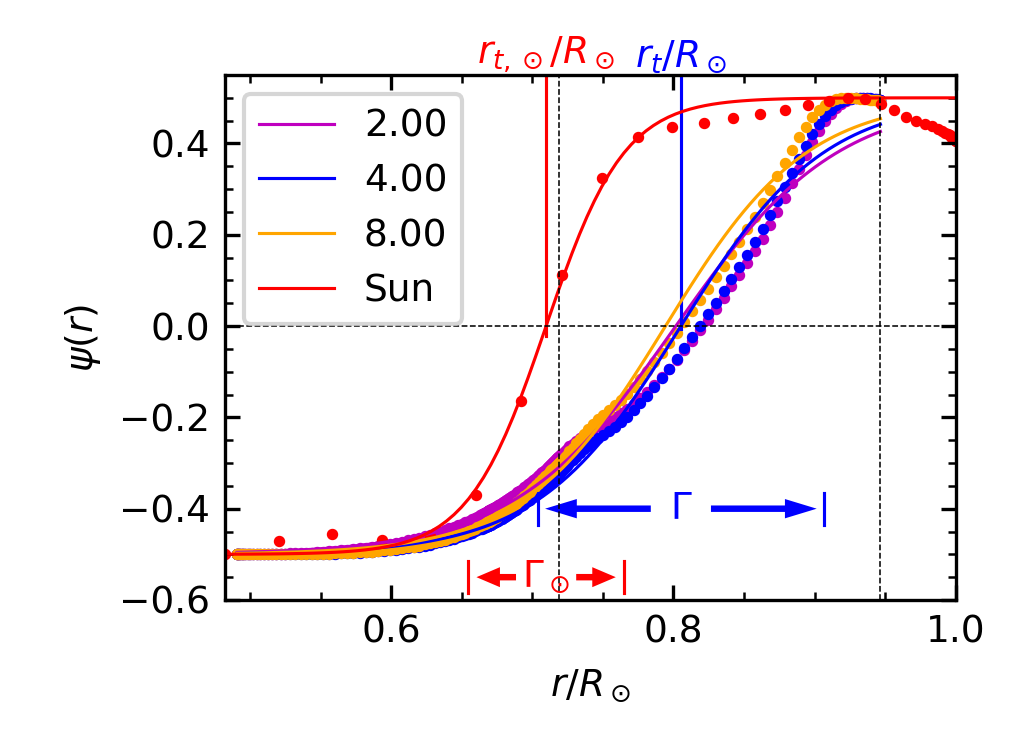}
	\caption{Scatter plots showing the $\psi$ profiles [Equation \eqref{eq:psi}] for a weak-, medium-, and strong-field case, as well as the Sun. Corresponding line plots show the function $(1/2)\tanh[2(r-r_t)/\Gamma]$ which is the best fit to $\psi$. Various lines and arrows indicate the values of $r_t$ and $\Gamma$ for Case 4.00 and the Sun.}
	\label{fig:drpsi}
\end{figure}

Figure \ref{fig:drpsi} shows the $\psi$ profiles for the Sun and some medium- and strong-field cases, along with the corresponding best-fit tanh functions. As we would expect, the solar $\psi$ (or best-fit tanh) profile has a centroid near $r=r_0$ and a relatively narrow width. For each simulated tachocline, the distributed radial shear in the CZ both widens the $\psi$ (or tanh) profile and pushes its centroid close the middle of the CZ. 

Table \ref{tab:tach} shows the tachocline parameters, as well as $\Delta\Omega\cz$, $\Delta\Omega\rz$, $f$, and $\Omega\rz$ for our simulations and the Sun. Clearly stronger fields reduce rotation contrast everywhere, but significantly more so in the RZ. Interestingly, the ``tachocline contrast ratio" $f$ is a non-monotonic function of regime and appears to be minimized (to a value of roughly 0.11) near $\prm=4$. The nominal solar value is quite a bit higher: $f_\odot\sim0.3$ from Table \ref{tab:tach}. However, this is mostly due to the uncertain tachocline width. The solar ``$\Delta\Omega\rz$" thus contains substantial contrast from the lower half of the tachocline. Deeper in the RZ $(r/\rsun\lesssim0.6)$ the helioseismic inversion gives $\Delta\Omega\sim0.014$ or $f_\odot\sim0.07$, which is probably closer to the value simulations should tend towards to be considered sufficiently solar-like.

In contrast to the Sun, all the simulated tachoclines have centroids well within the CZ and are roughly twice as thick. There is no clear scaling of the tachocline centroid with regime. However, the tachoclines in the strong-field regime are very slightly thinner than the tachoclines in the medium-field regime. Overall, it seems unlikely that further increasing the magnetic field strength will push the tachoclines to be more solar-like. Furthermore, there does not appear to be a solar-like ``sweet spot" (say, in between the medium- and weak-field regimes), wherein the RZ rotates like a solid body, but strong rotation contrast is sustained in the CZ. 

\begin{table*}
	\caption{Properties of the rotation rate for our simulations and the Sun. Properties for the weak-field cases are not shown, since they are almost identical to Case H. Note that for the Sun, we have $r_{t,\odot}/\rsun=0.72$ and $\Gamma_\odot/\rsun=0.11$, in reasonable agreement with the helioseismic estimates given in Section \ref{sec:intro}.} 
	\label{tab:tach}
	\centering
	\begin{tabular}{*{11}{l}  }
		\hline

Case   &   H   &   1.08   &   1.33   &   1.67   &   2.00   &   3.00   &   4.00   &   6.00   &   8.00   &   Sun  \\
regime   &   -   &   M   &     M   &   M   &   M   &   S   &   S   &   S   &   S  & -  \\
\hline
$\Delta\Omega_{\rm{CZ}}$   &   0.192   &   0.057   &   0.045   &   0.041   &   0.041   &   0.030   &   0.027   &   0.024   &   0.022   &   0.199  \\
$\Delta\Omega_{\rm{RZ}}$   &   0.116   &   0.016   &   0.011   &   9.13e-3   &   9.34e-3   &   3.34e-3   &   2.56e-3   &   2.57e-3   &   2.36e-3   &   0.046  \\
$f\definealt \Delta\Omega_{\rm{RZ}}/\Delta\Omega_{\rm{CZ}}$   &   0.603   &   0.287   &   0.239   &   0.221   &   0.227   &   0.112   &   0.094   &   0.106   &   0.108   &   0.232  \\
$\Omega_{\rm{RZ}}$   &   -0.025   &   -4.33e-3   &   -2.91e-3   &   -2.55e-3   &   -2.55e-3   &   -1.42e-3   &   -1.26e-3   &   -1.19e-3   &   -1.11e-3   &   -3.85e-3  \\
$r_t/R_\odot$   &   -   &   0.735 & 0.737 & 0.738 & 0.738 & 0.739 & 0.739 & 0.737 & 0.736 & 0.717   \\
$\Gamma/R_\odot$   &   -   &   0.247   &   0.234   &   0.226   &   0.229   &   0.209   &   0.203   &   0.199   &   0.201   &   0.111  \\

		\hline
	\end{tabular}
\end{table*}

\subsection{Torque Balance}\label{sec:torque}
In \citetalias{Matilsky2022}, we explicitly showed that the magnetic torque from the cycling, non-axisymmetric dynamo field was responsible for confining the tachocline in Case 4.00. This remains true for all the medium- and strong-field cases here. In the equilibrated state, the zonally and temporally averaged $\phi$-component of the momentum Equation \eqref{eq:mom} yields
\begin{subequations}	\label{eq:torque}
	\begin{align}
		&\underbrace {-\Div[\rhoref r \sin\theta \avphit{u^\prime_\phi \upol^\prime}]}_{\taurs \text{ (Reynolds stress)}} 
		-\underbrace{\rhoref\avt{ \avphi{\upol}\cdot\nabla\mathcal{L} } }_{\taumc \text{ (meridional circulation)}}+\nonumber\\
		&+\underbrace{\ek\nabla\cdot\left[\rhoref\nuref r^2\sin^2\theta\nabla\Omega\right]}_{\tauv \text{ (viscous)}}
		+  \underbrace{\nabla\cdot \left[r\sin\theta \avphit{ B_\phi^\prime\bpol^\prime}\right]}_{\taums \text{ (Maxwell stress)}}\nonumber\\
		&+  \underbrace{\nabla\cdot \left[r\sin\theta \avt{  \avphi{B_\phi}\avphi{\bpol} }\right]}_{\taumm \text{ (mean magnetic)}} = 0,\\
		&\where \underbrace{\mathcal{L}\definealt r\sin\theta(r\sin\theta + \avphi{u_\phi}) }_{\text{angular momentum density}},
	\end{align}
\end{subequations}
where for a vector field $\bm{A}$ we define its poloidal component $\bm{A}_{\rm pol}\definealt A_r\er + A_\theta\et$. The torques can be attributed to the physical processes labeled underneath each term (see \citealt{Miesch2011,Matilsky2019}). 
\begin{figure*}
	\centering
	\includegraphics{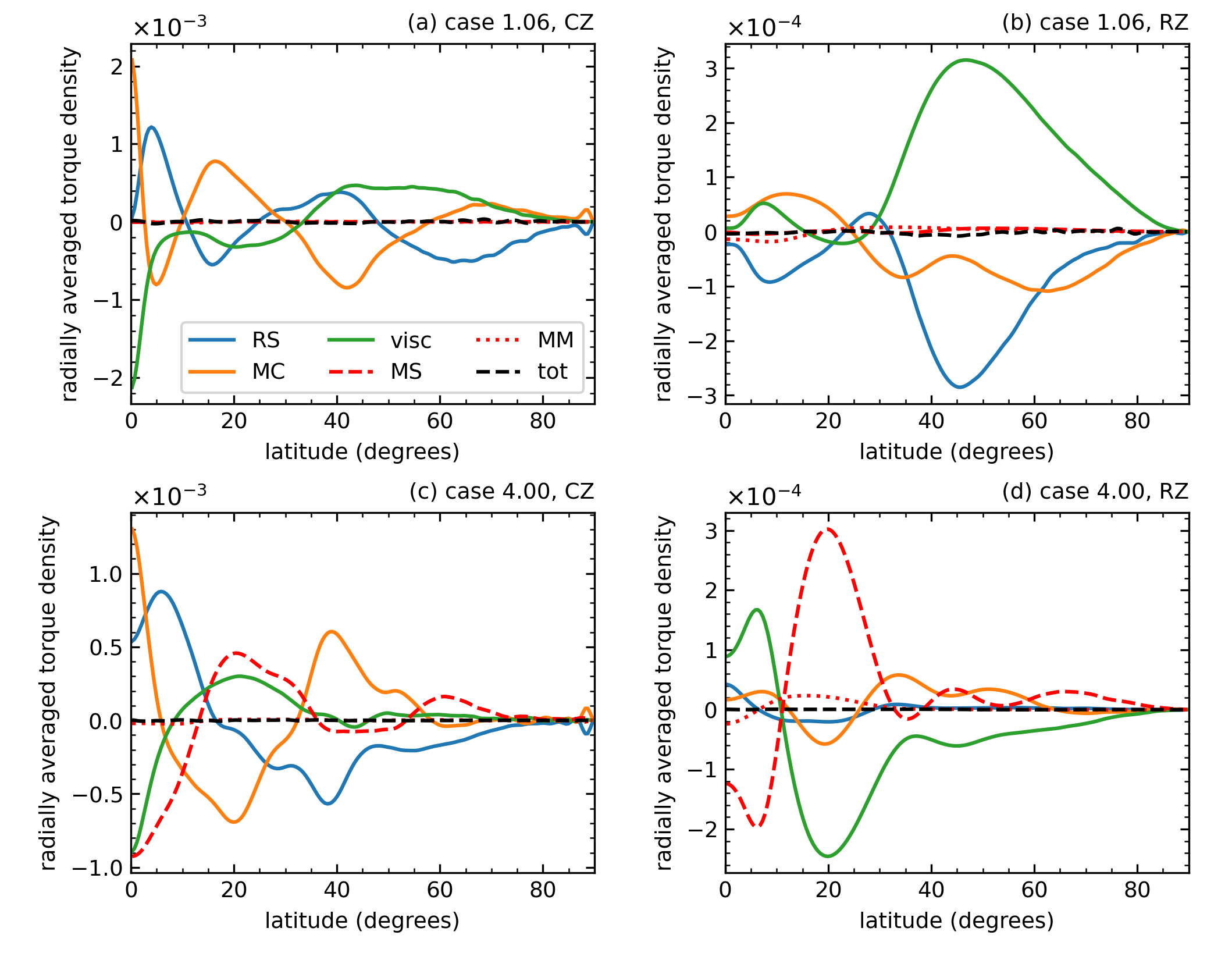}
	\caption{Torque densities, temporally averaged over the equilibrated state. The torques are also radially averaged, separately for the CZ (left-hand column) and RZ (right-hand column), and the equatorially symmetric parts are plotted as functions of latitude for the weak-field Case 1.06 (top row) and strong-field Case 4.00 (bottom row). The abbreviations in the legend shows which torque density from Equation \eqref{eq:torque} is plotted: Reynolds stress (RS), meridional circulation (MC), viscous (visc), Maxwell stress (MS), mean magnetic (MM), and total (tot).}
	\label{fig:torque}
\end{figure*}

Figure \ref{fig:torque} shows the full steady-state torque balance (in the CZ and RZ separately) for Case 1.06 (the strongest weak-field case) and the strong-field Case 4.00. The CZ of Case 1.06 [Figure \ref{fig:torque}(a)] is effectively hydrodynamic in its torque balance. It represents the ``standard" by which current global models (e.g., \citealt{Hotta2015,Guerrero2016a,Matilsky2019}) maintain a solar-like  differential rotation with fast equator and slow pole. That is, the rotational influence on the convection lead to Taylor columns with correlations in the components of $\vecu^\prime$ (i.e., Reynolds stresses), which transport angular momentum away from the rotation axis and produce mostly positive torques at low latitudes ($\lesssim45^\circ$; further from the rotation axis) and negative torques at high latitudes ($\gtrsim45^\circ$; closer to the rotation axis). Meridional circulation also plays a role (a complicated one due to the presence of multiple circulation cells), especially at low latitudes. Viscosity always tries to eliminate gradients in $\Omega$, in this case the latitudinal gradients, by spinning the equator down and the polar regions up. \newtext{Note that this downward viscous spread of differential rotation is distinct from spread along poloidal field lines according to Ferraro's law (e.g., \citealt{Strugarek2011a,Strugarek2011b}) and in general, the RZ's isorotation contours in the weak-field cases do not fall along poloidal field lines.} The RZ of Case 1.06 [Figure \ref{fig:torque}(b)] has a torque balance that is roughly an imprint of the balance in the CZ, but is overall much weaker and concentrated at high latitudes (and the magnetic torques are negligible). 

The torque balance in the CZ of Case 4.00 [Figure \ref{fig:torque}(c)] still has a positive Reynolds-stress torque, but this positive torque is confined to significantly lower latitudes ($\lesssim15^\circ$). Consequently, most differential rotation is confined to a narrow prograde jet at the equator. The Maxwell-stress torque opposes the Reynolds-stress torque and effectively acts as an additional source of viscous torque. The meridional-circulation torque is significantly altered from its weak-field counterpart as well. Evidently, strong-field magnetism not only provides an additional torque, but also changes the structure of the convection and circulation so as to alter the hydrodynamic torques from their weak-field forms.

Finally, the torque balance in the RZ of Case 4.00 [Figure \ref{fig:torque}(d)] was studied in \citetalias{Matilsky2022} and clearly this is the balance responsible for tachocline confinement. The profile of viscous torque has changed sign compared to the torque profiles in the other panels: it is now positive at low latitudes ($\lesssim15^\circ$) and negative at high latitudes ($\gtrsim15^\circ$), thus trying to imprint the equatorial jet and weak high-latitude retrograde differential rotation downward.\footnote{Note that the viscous torque always attempts to eliminate gradients in $\Omega$; however, in the presence of other torques, it cannot eliminate the gradient in all directions. In the case of a radial shear layer like the tachocline, viscosity will reduce $|\pderivline{\Omega}{r}|$ at the expense of imprinting the latitudinal differential rotation downward, which of course increases $|\pderivline{\Omega}{\theta}|$.} The viscous torque is countered by the magnetic torque, which must come from the large-scale, non-axisymmetric ($m=1,2$) field components shown in Figure \ref{fig:sslice} (that this is true, at least for Case 4.00, was shown explicitly in \citetalias{Matilsky2022}). 

All our weak- and strong-field cases have torque balances like those in Figure \ref{fig:torque}. The medium-field cases have balances essentially similar to the strong-field cases, but the torques become more complicated due to the intermittent changes in field strength and differential rotation that was noted in connection with Figure \ref{fig:energy}. Regardless, the answer to how our simulated tachoclines are confined reduces to explaining the maintenance of large-scale, non-axisymmetric magnetism in the RZ. The following sections show how this maintenance can be understood in terms of the cycling dynamo and skin effect. 

\begin{table*}
	\caption{Dynamo cycle properties for each magnetic case ($\omcyc$, $\sigma_\omega$, $\pcyc$, and $q$), as defined in Equation \eqref{eq:omcyc}. Here, $\delta t$ and $\sigma_t$ are the mean and standard deviation in the sample rate for the spherical-slice magnetic field data, $\omnyq\definealt 2\pi/(2\delta t)$ is the (angular) Nyquist frequency, and $\delta\omega\definealt 2\pi/(\tmax-\tequil)$ is the (angular) frequency resolution.} 
	\label{tab:cyc}
	\centering
	\begin{tabular}{*{13}{l}  }
		\hline

Case   &   1.00   &   1.05   &   1.06   &   1.08   &   1.33   &   1.67   &   2.00   &   3.00   &   4.00   &   6.00   &   8.00  \\
Regime   &   W   &   W   &   W   &   M   &   W   &   M   &   M   &   M   &   S   &   S   &   S  \\
\hline
$\omcyc$   &   6.07e-4   &   5.80e-4   &   5.20e-4   &   -5.21e-3   &   1.56e-3   &   -4.65e-3   &   -5.13e-3   &   -2.52e-3   &   -1.74e-3   &   -1.45e-3   &   -1.52e-3  \\
$\pcyc/\prot$  &    1648   &    1724   &    1923   &   191.9   &   642.2   &   215.1   &   194.9   &   396.1   &   574.1   &   692.0   &   656.9  \\
$\sigma_\omega$   &   9.97e-5   &   1.62e-4   &   2.89e-4   &   3.52e-3   &   2.47e-3   &   7.35e-3   &   0.011   &   3.71e-3   &   1.74e-3   &   1.86e-3   &   9.51e-4  \\
$q\definealt \omcyc/\sigma_\omega$   &   6.09   &   3.57   &   1.80   &   1.48   &   0.63   &   0.63   &   0.46   &   0.68   &   1.00   &   0.78   &   1.60  \\
\hline
$\delta t/\prot$   &   3.76   &   3.79   &   3.76   &   4.10   &   4.08   &   4.03   &   3.95   &   3.53   &   3.18   &   2.87   &   2.65  \\
$\sigma_t/\prot$   &   0.40   &   0.37   &   0.37   &   0.08   &   0.08   &   0.14   &   0.21   &   0.30   &   0.25   &   0.17   &   0.14  \\
$\omnyq$   &   0.133   &   0.132   &   0.133   &   0.122   &   0.122   &   0.124   &   0.127   &   0.141   &   0.157   &   0.174   &   0.189  \\
$\delta\omega$   &   1.21e-4   &   1.93e-4   &   1.73e-4   &   1.41e-4   &   1.73e-4   &   1.50e-4   &   1.25e-4   &   1.49e-4   &   6.97e-5   &   2.06e-4   &   1.90e-4  \\

		\hline
		
	\end{tabular}
\end{table*}

\section{Cycling Behavior}\label{sec:cycle}
\subsection{Dynamo Cycles in the Weak- and Strong-Field Regimes}\label{sec:cycle1}
\begin{figure*}
	\centering
	\includegraphics[width=7.25in]{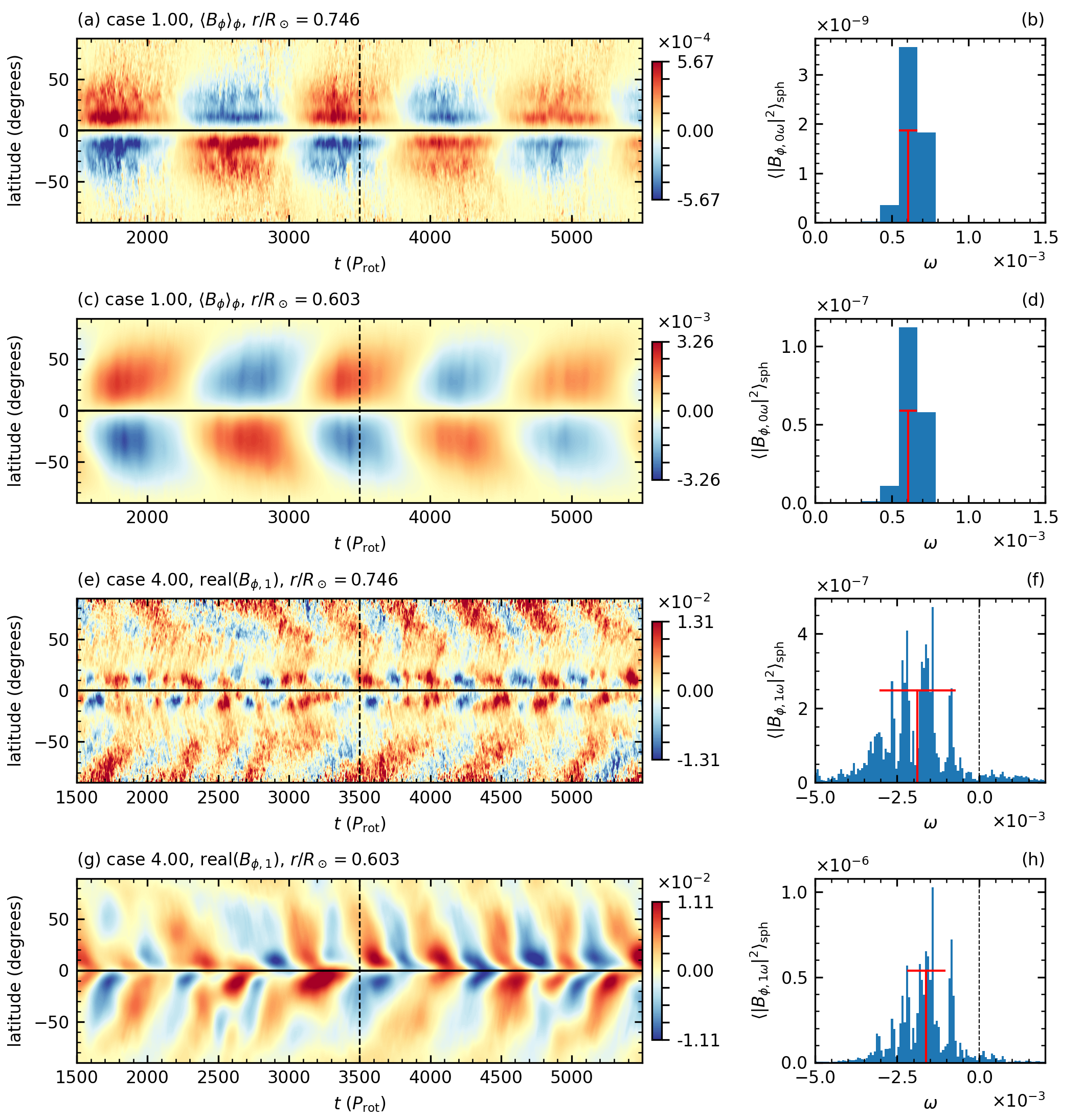}
	\caption{Time-latitude diagrams of the large-scale ($m=0$ or $1$) toroidal field over the interval $(1500,5500)\prot$ for a weak-field, axisymmetric dynamo (Case 1.00; upper 4 panels) and a strong-field, non-axisymmetric dynamo (Case 4.00; lower 4 panels). For each case, we sample the same two depths as Figure \ref{fig:sslice}. In each time-latitude diagram, the horizontal solid line marks the equator and the vertical dashed line marks $t=3500$ $\prot$, the instant sampled by Figure \ref{fig:sslice}. To the right of each time-latitude diagram, we show (for the same depth and $m$-value as the time-latitude plot) the latitudinally averaged toroidal-field powerspectrum $P(\omega)=\avsph{|B_{\phi,m\omega}|^2}$ [see Equation \eqref{eq:bmw}; here, $m$ is 0 or 1]. Since $\vecb_0=\avphi{\vecb}$ is real, we consider $P(\omega)$ a function of positive $\omega$ only when $m=0$. The red ``T" marks the location of the primary cycle frequency $\omcyc$ and the dispersion $\sigma_\omega$ for $P(\omega)$ [see Equation \eqref{eq:omcyc}]. For Case 4.00 (panels f,h), $\omega=0$ is marked by a vertical dashed line. }
	\label{fig:timelat}
\end{figure*}

Figure \ref{fig:timelat} (left-hand panels) shows time-latitude diagrams of $\avphi{B_\phi}$ for the weak-field Case 1.00 and $\text{real}(B_{\phi,1})$ for the strong-field Case 4.00 at two depths, one in the CZ and one in the RZ. Both cases cycle, although the polarity reversals in the weak-field case occur significantly more regularly than the reversals in the strong-field case. In each case, the cycle ``imprints" from the base of the CZ onto the RZ with a phase lag (i.e., for every reversal in the CZ, there is a corresponding reversal in the RZ some time later). There is also significantly more rapid variation in the large-scale field in the CZ (seen as graininess in the time-latitude plots) than in the RZ. This again suggests that the RZ acts as a low-pass filter, in time as well as in space. 

To describe these cycles more precisely, we define the frequency components of each $\vecb_m$:
\begin{align}
	\vecb_{m\omega} &\definealt \avt{\vecb_m e^{i\omega t}W(t)} = \avphit{\vecb e^{-i(m\phi-\omega t)}W(t)},\label{eq:bmw}
\end{align}

where $W(t)$ is the Hanning window function and $\omega$ is the discrete angular frequency. From the convention in the exponential (for nonzero $m$ only), the $\vecb_{m\omega}$ components with positive $\omega/m$ move prograde in longitude and the components with negative $\omega/m$ move retrograde. 

We sample the spherical-slice magnetic-field data during the equilibrated state ($\tequil$ to $\tmax$). The sampling intervals are not uniform within a given simulation, but they are typically close to the mean interval $\delta t\sim3$--$4 \prot$, with a typical standard deviation of $\sigma_t\sim0.1$--$0.4\prot$ (see Table \ref{tab:cyc}). We thus interpolate the non-uniform time series onto a uniform time series spaced by $\delta t$ before computing the (windowed) discrete Fourier transform represented by Equation \eqref{eq:bmw}. 

Figure \ref{fig:timelat} (right-hand panels) shows the power in the large-scale toroidal field ($|B_{\phi,0\omega}|^2$ for the weak-field case and $|B_{\phi,1\omega}|^2$ for the strong-field case) corresponding to the time-latitude diagrams. The regularity of the weak-field cycle causes most of the power to be concentrated in the primary central frequency. By contrast, for the irregular strong-field cycle, there is a wide dispersion of power around a negative central frequency. This preference for negative frequencies suggests retrograde propagation of $\vecb_1$, broadly consistent with transport by the negative background rotation rate in the RZ. Furthermore, the high-$|\omega|$ ``tail" in the strong-field case is significantly less pronounced in the RZ than in the CZ, again reinforcing the idea that the RZ acts as a low-pass filter in time.

Figure \ref{fig:timelat} (right-hand panels) shows that in each case, there is a central frequency (the ``primary" cycle frequency $\omcyc$) and a dispersion (of width $\sigma_\omega$) in power about this central frequency. More precisely, for a given powerspectrum $P(\omega)$, we define $\omcyc$ as the median frequency associated with $P(\omega)$ and $\sigma_\omega$ as $P(\omega)$'s half-integral width:
\begin{align}
	\sum_{\omega\leq\omcyc}P(\omega)=  \sum_{\omega=\omcyc-\sigma_\omega/2}^{\omcyc+\sigma_\omega/2}   P(\omega)\definealt \frac{1}{2} \sum_\omega P(\omega).\label{eq:omcyc}
\end{align}
The cycle period is $\pcyc\definealt2\pi/\omcyc$ (since $\prot=2\pi$, note that $\pcyc/\prot=1/\omcyc$). The quantity $q\definealt\omcyc/\sigma_\omega$ defines the regularity of the cycle, with higher $q$ indicating a more regular cycle.

Table \ref{tab:cyc} contains values of $\omcyc$, $\pcyc$, $\sigma_\omega$, and $q$, along with the sampling parameters $\delta t$, Nyquist frequency $\omnyq$, and frequency resolution $\delta\omega$. For the weak-field cases, we take $P(\omega)= \av{|\vecb_{0\omega}|^2}\full$ (considering positive $\omega$ only, since $\vecb_0=\avphi{\vecb}$ is real) and for the non-weak-field cases, we take $P(\omega)= \av{|\vecb_{1\omega}|^2}\full$ (considering both positive and negative $\omega$). The weak-field solutions all have similar cycle periods ($\pcyc\sim1400$--$2000\prot$), with relatively high values of $q$. This confirms the visual appearance of regular cycles in the weak-field cases (Figure \ref{fig:timelat}). 

The medium- and strong-field cases have more irregular cycles (with $q\lesssim1$) and the cycle period (with the exception of either one of cases 1.67 or 2.00) monotonically increases with increasing field strength. Since field strength increases with $\prm$ and therefore with magnetic diffusion time $P_\eta$ (see Table \ref{tab:inputnondimderiv}), this suggests that the cycle period for the non-weak cases is at least partly determined by the level of diffusion (i.e., $\pcyc$ scales more or less monotonically with $P_\eta$). 

\section{Skin-Depth Interpretation}\label{sec:skindepth}
As mentioned at the conclusion of Section \ref{sec:torque}, explaining the presence (or not) of tachoclines in these simulations boils down to the maintenance of large-scale, non-axisymmetric ($m=1,2$) $\bpol$ in the RZ.\footnote{Maintenance of large-scale $B_\phi$ is also important of course. However, if $\bpol$ is present, $B_\phi$ is always created by mean shear. In \citetalias{Matilsky2022}, we argued that this effect---similar in essence to Ferraro's law \citep{Ferraro1937}---is in fact responsible for the magnetic torque and hence tachocline confinement. In this work, we thus only consider the maintenance of $\bpol$.} In \citetalias{Matilsky2022}, we showed that two effects were responsible for this maintenance: induction (possibly from inertial oscillations; see also \citealt{Blume2024}) and diffusion of CZ-produced field to roughly a (then ill-defined) skin-depth below the CZ. In this section, we precisely define the relevant skin effect and we show how the amplitude of $\bpol$ in the RZ can be extremely well-predicted considering only diffusive skin effects.  

\begin{figure}
	\centering
	\includegraphics{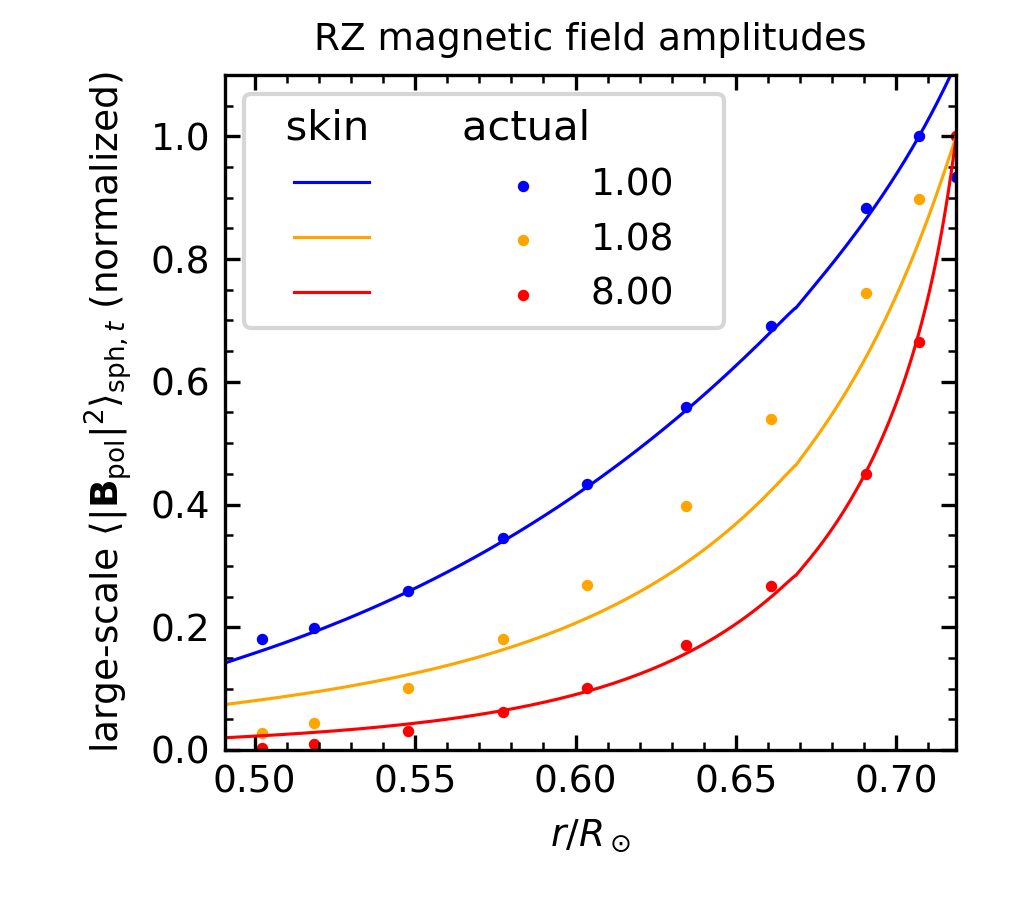}
	\caption{Amplitude (in the RZ) of ``large-scale $\avspht{|\bpol|^2}$", defined here as $\sum_m\avspht{|\vecb_{{\rm pol},m}|^2}(r)$, where the sum is over $m=0$ for weak-field cases and $m\in \{0,1,2\}$ for medium- and strong-field cases. We show both the actual amplitude (solid dots) and the amplitude predicted by the skin-depth Equation \eqref{eq:ampskin} (solid curves) for Cases 1.00, 1.08, and 8.00. For Case 1.00, we replace $r_0$ in Equation \eqref{eq:ampskin} with a value $r_c$ slightly below the CZ: $r_c/\rsun=0.707$. Each profile is normalized such that its value at $r=r_0$ (or $r=r_c$ for Case 1.00) is unity.}
	\label{fig:skin}
\end{figure}

As a first approximation, we assume that fluid motions produce no electromotive force (e.m.f.) below $r_0$ (or a radius slightly below $r_0$ for the weak-field cases; see Figure \ref{fig:skin}'s caption). Then the evolution of $\bpol$ in the RZ is governed by diffusion alone, with the upper boundary condition (at $r=r_0$) that $\bpol$ matches what the CZ produces and the lower boundary condition (at $r=r\inn$) that the field decays with depth instead of grows. For axisymmetric weak-field dynamos, the regular polarity reversals provide an oscillating boundary condition at a single frequency. This is the classic form of Stokes' problem of an oscillating boundary. In its solution, the field amplitude is contained in an envelope that decays exponentially downwards with a scale height (in this context, called the skin depth) that depends on the frequency of oscillation. This is the formalism expounded in the original fast magnetic confinement scenario of \citet{ForgcsDajka2001}. Note that in this axisymmetric case, the rotation rate of the frame in which the equations are solved does not matter. 

However, for the non-axisymmetric medium- and strong-field dynamos, the choice of rotating frame does matter. Since advection in $\phi$ of a non-axisymmetric $\bpol$ constitutes an e.m.f., diffusion-only evolution is possible only if the RZ rotates approximately like a solid body. Then, to examine purely diffusive solutions, the induction equation must be written in the frame rotating at the solid-body rate $\Omega\rz$ (see Table \ref{tab:tach} for the simulated and solar values of $\Omega\rz$). Because the field at $r=r_0$ is cycling with multiple frequencies (see the previous Section \ref{sec:cycle}), this setup still corresponds to Stokes' problem, but there is now a different skin-depth for each component $\vecb_{{\rm pol},m\omega}$. Furthermore, since the equations must be solved in the frame of the RZ, the frequency determining the skin-depth is not $\omega$, but the Doppler-shifted value $\omega-m\Omega\rz$.\footnote{Note that the relative signs of $\omega$ and $\Omega\rz$ matter here, but the sign of $\omega-m\Omega\rz$ does not; see Equation \eqref{eq:skindepthmw}.} 

Assuming that the spatial variation of $\bpol$ is predominantly radial, Equation \eqref{eq:ind} leads to separate boundary-value problems for each $\vecb_{{\rm pol},m\omega}$:
\begin{align}\label{eq:indskin}
	-i(\omega-m\Omega\rz)\vecb_{{\rm pol},m\omega} &\approx \frac{\ek}{\prm}\etaref(r)\pderiv{^2\vecb_{{\rm pol},m\omega}}{r^2}
\end{align}
for $r\leq r_0$. Rapid variation in $r$ allows us to neglect the terms in $\nabla^2$ other than $(\pderivline{}{r})^2$, sphericity terms, and the term from $\nabla\etaref$. Note that Equation \eqref{eq:indskin} is valid for all $m$.

Because $\etaref(r)$ varies with radius, we follow \citet{Garaud1999} and define
\begin{align}\label{eq:reta}
	r_\eta \definealt r\inn + \frac{\int_{r\inn}^{r}\etaref(r^\prime)^{-1/2}dr^\prime}{\int_{r\inn}^{r_0}\etaref(r^\prime)^{-1/2}dr^\prime}.
\end{align}
Note that $r_\eta$ is a monotonically increasing function of $r$ and is equal to $r$ at $r=r\inn$ and $r=r_0$.\footnote{We believe that the $r_\eta$ given in \citet{Garaud1999}, which had $\etaref(r^\prime)^{+1/2}$ in the integrand in the analog of Equation \eqref{eq:reta}, was mistakenly defined.}  Again assuming rapid radial variation, Equation \eqref{eq:indskin} becomes
\begin{subequations}\label{eq:indskin2}
	\begin{align}
	-i(\omega-m\Omega\rz)\vecb_{{\rm pol},m\omega} \approx \frac{\ek}{\prm}\etaref\const \pderiv{^2\vecb_{{\rm pol},m\omega}}{r_\eta^2},\\
		\where \etaref\const \definealt \left[ \frac{r_0-r\inn}{\int_{r\inn}^{r_0}\etaref(r^\prime)^{-1/2}dr^\prime} \right]^2
	\end{align}
\end{subequations}
is an intermediate value of $\etaref(r)$ in the RZ. For our chosen reference state, $\etaref\const=0.292$ and $\etaref\ofr$ achieves this value at $r/\rsun=0.599$.

Equation \eqref{eq:indskin2} is of course Stokes' problem again and its exact solution yields
\begin{subequations}\label{eq:ampskin}
\begin{align}
	\avsph{|\vecb_{{\rm pol},m\omega}|^2}(r) = &\avsph{|\vecb_{{\rm pol},m\omega}|^2}(r_0)\times\nonumber\\
	& \exp{\left[-2\left(\frac{r_0-r_\eta}{\delta_{m\omega}} \right)\right]},\\
	\where \delta_{m\omega} \definealt & \ \sqrt{\frac{2\ek\etaref\const}{\prm|\omega-m\Omega\rz|} }\label{eq:skindepthmw}
\end{align}
\end{subequations}
is the $m$- and $\omega$-dependent skin-depth. 

The ``skin-predicted" amplitude of large-scale $\avspht{|\bpol|^2}$ is then found by summing Equation \eqref{eq:ampskin} over all $\omega$ and low $m$. We choose $m=0$ for the weak-field cases and $m\in\{0,1,2\}$ for the medium- and strong-field cases. Figure \ref{fig:skin} shows large-scale $\avspht{|\bpol|^2}$ (both the skin-predicted and actually-realized values) for a weak-, medium-, and strong-field case. Equation \eqref{eq:ampskin} does an extremely good job of predicting the field strength for the weak- and strong-field cases and a reasonable job for the medium-field case. Overall, it thus seems highly likely that the magnetization of the RZ is determined primarily by the dynamo cycle of the CZ imprinting diffusively downward. 

In \citetalias{Matilsky2022}, the strong $\bpol$ in the RZ of Case 4.00 was attributed partially to deep dynamo action. For all the magnetic cases considered here, we have verified that the deep dynamo is still present, that is, the production of $|\bpol|^2$ by diffusion ($D_{\rm pol}$) is negative in the RZ, while the production by e.m.f. ($I_{\rm pol}$) is positive.\footnote{Explicitly, we define $D_{\rm pol}\ofr\definealt \avspht{\bpol\cdot[\curl (\etaref\curl\vecb)]_{\rm pol}}$ and $I_{\rm pol}\ofr\definealt \avspht{\bpol\cdot[\curl (\vecu\times\vecb)]_{\rm pol}}$. We have verified that in all magnetic cases, at all radii in the RZ,  $D_{\rm pol}(r)<0$, while $I_{\rm pol}(r)>0$.} It was emphasized in \citetalias{Matilsky2022} that this implies (by definition; e.g., \citealt{Moffatt2019}, p. 146) the presence of dynamo action deep in the RZ, and we argued in \citetalias{Matilsky2022} that this deep dynamo (possibly driven by Rossby waves) may have been responsible for tachocline confinement in Case 4.00. However, the results of this section indicate that the strength of $\bpol$ in the medium- and strong-field cases can be almost fully accounted for by diffusive skin effects. It thus seems likely we would have tachocline confinement (in the simulations considered here) regardless of whether there was a deep dynamo or not. How the deep dynamo is driven---and whether it \textit{can} confine the tachocline in the absence of large diffusion---remains an intriguing open question.

\section{Polarity Reversals for Non-Axisymmetric Magnetic Fields}\label{sec:cyclena}
Polarity reversals in non-axisymmetric magnetic fields [e.g., Figures \ref{fig:timelat}(e,g)] can be accomplished in two distinct ways. For definiteness, consider $B_{\phi,1}$ (i.e., the colatitudinal field associated with a single partial-wreath pair). At a given radius and latitude, we have 
\begin{align}\label{eq:tworeversals}
	W(t)B_{\phi,1}(t) = \sum_\omega B_{\phi,1\omega}e^{-i\omega t} = A(t)e^{i\varphi(t)}
\end{align}
\newtext{The first equality comes directly from Equation \eqref{eq:bmw} and the second equality is simply the mathematical statement that any complex number can be written as an amplitude [here $A(t)$] and a complex phase [here $e^{i\varphi(t)}$]. Note that the presence or not of the window function $W(t)$ is immaterial to the following arguments.}

The first type of non-axisymmetric polarity reversal is due to modulation of the amplitude $A(t)$. These reversals contain cycle minima [for which $A(t)=0$] and are analagous to the reversals of full-wreath (i.e., axisymmetric) polarity in the weak-field cases [e.g., Figures \ref{fig:timelat}(a,c)], or equivalently to what we believe happens to the solar interior magnetic field to cause the observed butterfly diagram (e.g., \citealt{Hathaway2015}). The second type of non-axisymmetric polarity reversal is due to changes in the phase $\varphi(t)$, which simply occur from advection of the whole structure in longitude (there are no cycle minima in this case). Equation \eqref{eq:tworeversals} shows that in general, there is no straightforward way to separate which frequency components $B_{\phi,1\omega}$ are due to each type of reversal. Indeed, in \citet{Matilsky2020a} (see Figures 11 and 12 from that paper), we showed that both types of reversal occur simultaneously in the CZ-only partial-wreath cycles, with the frequency of amplitude modulation similar to that of longitudinal advection.

\begin{figure*}
	\centering
	\includegraphics[width=7.25in]{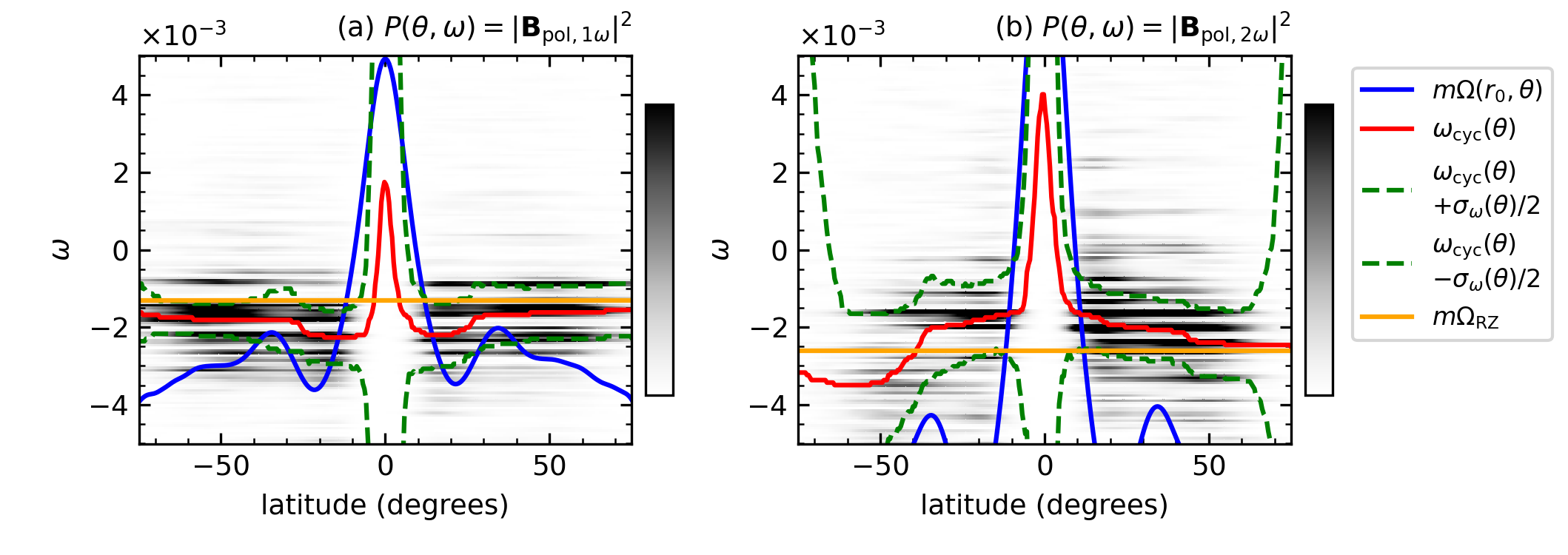}
	\caption{Powerspectra $P(\theta,\omega)$ of the poloidal field at $r=r_0$ in Case 4.00 (viewed as functions of latitude and frequency) for (a) $m=1$ and (b) $m=2$. Power is shown in gray-scale in arbitrary units (a linear color-scaling is used, with white corresponding to the zero point). Overplotted is the local advective rotation rate $m\Omega(r_0,\theta)$, the location of most of the power [i.e., the $\theta$-dependent values $\omcyc(\theta)$ and $\omcyc(\theta)\pm\sigma_\omega(\theta)/2$; see Equation \eqref{eq:omcyc}], and the advective rotation rate of field in the RZ, $m\Omega\rz$.}
	\label{fig:pspecdr}
\end{figure*}

Postponing for now the important investigation of how amplitude modulation occurs (it must be caused by non-axisymmetric dynamo processes; e.g., \citealt{Stix1971,Ivanova1985,Bigazzi2004,Moss2002}), we discuss in this section which frequencies are consistent with longitudinal advection. Figure \ref{fig:pspecdr} shows Case 4.00's poloidal powerspectra as functions of latitude and frequency for both $m=1$ and $m=2$ at the CZ--RZ interface $r=r_0$. For $m=1$ (panel a), the shape of the powerspectrum is nearly latitude-independent, with roughly constant values of the latitudinally dependent cycle frequency $\omcyc(\theta)$ and dispersion $\sigma_\omega(\theta)$. The central frequency $\omcyc(\theta)$ overlaps with $m\Omega(r_0,\theta)$ at low latitudes (about $15^\circ$ north and south), which correspond to the retrograde ($\Omega<0$) jet of Figure \ref{fig:dr}(c). For $m=2$ (panel b), $\omcyc(\theta)=m\Omega(r_0,\theta)$ at a slightly lower latitude (about $10^\circ$ north and south) and the dispersion of power $\sigma_\omega(\theta)$ is ``stretched" by roughly a factor of two compared to the dispersion for $m=1$. This stretching factor is consistent with the advective rate being proportional to $m$-value. 

One valid interpretation of Figure \ref{fig:pspecdr} is that the large-scale ($m=1,2$) structures move as a cohesive structure (i.e., with rotation rates more or less independent of latitude), but at a time-varying rotation rate, which corresponds to a range of frequencies $\omcyc\pm\sigma_\omega/2$ centered about the advective rate $m\Omega(r_0,\theta)$ at low latitudes. Since these low latitudes are also the location of the retrograde jet, we may interpret the partial wreaths as being ``anchored" to the jet. Another valid interpretation is that the partial wreaths have an intrinsic rate of rotation caused by the dynamo mechanism, which is apparently independent of latitude. This dynamo-intrinsic rotation rate then \textit{determines} the rotation rate of the retrograde jet via the magnetic torques (since the magnetic torque should force the fluid to move with $\bpol$). 

For the skin depth [Equation \eqref{eq:skindepthmw}], it does not matter which physical process produces a given value of $\omega$. As long as $\omega$ is different from $m\Omega\rz$ (see the orange lines in Figure \ref{fig:pspecdr}), the amplitude of $\bpol$ should decay downward with a finite skin-depth. Considering the first interpretation of Figure \ref{fig:pspecdr}, we thus argue for the relevance of an important new type of skin effect \newtext{that could operate in stars with both rigidly-rotating RZs and non-axisymmetric magnetic fields.} This skin effect would arise from non-axisymmetric field at the CZ--RZ interface being advected by a background rotation rate that is different from the rotation rate of the RZ. 

One particularly interesting consideration are the latitudes at the base of the CZ that co-rotate with the RZ. At those latitudes, $\omega-m\Omega\rz=0$ and the skin depth in Equation \eqref{eq:skindepthmw} becomes infinite. What this really means is that any frozen-in non-axisymmetric field appears completely stationary to the RZ and spreads downward indefinitely on a diffusive time-scale. We explore this idea in the solar context in the following section.

\section{Discussion: Non-Axisymmetric Dynamo Confinement of the Solar Tachocline}\label{sec:concl}
These results suggest that the fast magnetic confinement scenario---which was originally proposed, in 1D only, for axisymmetric $\bpol$ cycling at a single frequency \citep{ForgcsDajka2001,ForgcsDajka2002,ForgcsDajka2004,Barnabe2017}---should be expanded (into 3D) to include both non-axisymmetric $\bpol$ and a spread in cycle frequencies. Whether this more general scenario is actually capable of confining the solar tachocline depends on several major differences between simulations and the Sun, which we briefly discuss here. Note that for this section (which is concerned with a real astrophysical object, namely, the Sun), we regard all physical quantities as dimensional.

Prior work on the fast magnetic confinement scenario has always assumed a turbulently enhanced magnetic diffusivity. For example, \citealt{Barnabe2017} (see Figure 5 from that paper) nicely show that for dynamo poloidal field strengths of $\sim$$10^3$ G (and a cycle period of $\sim$22 yr), the magnetic diffusivity must be larger than its molecular value by a factor of at least $10^5$--$10^6$. However, as discussed in Section \ref{sec:intro}, how much turbulent enhancement of the viscosity occurs in the hydrodynamic scenario (and if the enhancement is primarily horizontal or vertical) is a subject of ongoing research with no firm conclusions at present. For \textit{magnetic} diffusive enhancement, even less is known. Thus for simplicity, we assume here that the magnetic diffusion is not turbulently enhanced. We also leave aside for now \citetalias{Matilsky2022}'s proposition that deep dynamo action may generate significant $\bpol$. 

\subsection{Diffusive Equilibration in Simulations}
In the Sun, all diffusive time-scales are \newtext{significantly} greater than the current solar age ($t_\odot=4.6$ Gyr; see Table \ref{tab:solaranalog}). By contrast, simulations that seek to address the tachocline confinement problem do so by evolving the MHD equations over significant fractions of the relevant diffusion times. If a statistically steady state is achieved, it thus likely contains significant diffusive effects in the dynamical balances.

We believe this may be one of the main reasons our tachocline cases have most of the differential rotation confined to a narrow equatorial jet near the outer boundary. The viscous and magnetic diffusion time-scales are similar (we have order-unity $\prm$ values) and in most cases we run for several of each time-scale. The steady state thus necessarily has similar magnitudes for the viscous and magnetic torques in the CZ and RZ (compare the left-hand and right-hand columns of Figure \ref{fig:torque}). This means that any magnetic torque strong enough to prevent viscous tachocline spread is also strong enough to eliminate much of the differential rotation in the CZ.

\subsection{Viscous versus Radiative Spread: General Torque Balance}
Even barring the open question of whether circulation burrowing is hyperdiffusive in the Sun, it seems likely that radiative spread dominates viscous spread. This dominance is expressed via the  ``$\sigma$-parameter" (e.g., \citealt{Garaud2008a,Garaud2009,Wood2012,AcevedoArreguin2013,Wood2018}):
\begin{align}\label{eq:sigma}
	\sigma\definealt \sqrt{\frac{\pes}{\pnu_{,\rm RZ}}}=\frac{\sqrt{\pr\rm Bu}}{2}
\end{align}
 For the Sun, $\sigma_\odot=0.17\ll1$ (see Table \ref{tab:solaranalog}). Since the Reynolds number in the solar CZ is extremely high, the viscous torque should drop out of the torque balance in the CZ as well. Global simulations seem to indicate that large-scale magnetic field (when strong enough) significantly reduces the differential rotation in the CZ (e.g., \citealt{Brown2010,Racine2011,Passos2014,Yadav2015,Augustson2015,Guerrero2019,Bice2020,Matilsky2020a,Matilsky2020c}). For a fast magnetic confinement scenario to work (i.e., a scenario in which the dynamo-produced magnetic field diffusively penetrates into the upper RZ), we thus might require that the total magnetic torque ($\taumag\definealt \taums + \taumm$) be both large enough in the RZ to counter radiative spread and small enough to drop out of the torque balance in the CZ. In that case, Equation \eqref{eq:torque} (its dimensional counterpart) becomes 
\begin{align}\label{eq:torquegeneral}
	0=\begin{cases}
		&\underbrace{-\frac{4\omsun^2}{\overline{N^2}}r_0^2\rhoref\ \kapparef\pderiv{^4\avt{\mathcal{L}}}{r^4}}_{\taurad \text{ (radiative spread)}} + \taumag\five \text{in the RZ}\\
		&\taurs + \taumc\five\five\five\five\five\ \ \  \text{in the CZ,}
	\end{cases}
\end{align}
where the form of $\taurad$ is derived in \citet{Spiegel1992} [their Equation (4.9)] and we have assumed a thin tachocline (so that we retain only highest derivatives in $r$). We estimate $\pderivline{^4\avt{\mathcal{L}}}{r^4}\sim (r_0/\sqrt{2})^2\Delta\Omega\cz/\Gamma_\odot^4$. If we take $\Gamma_\odot=0.05\rsun$ and take the Model S values in Table \ref{tab:solaranalog} (averaged over the upper solar RZ) for $\rhoref$, $\kapparef$, and $\overline{N^2}$, and take $\Delta\Omega\cz=0.20\omsun$ from Table \ref{tab:tach}, we find 
\begin{align}\label{eq:tauradest}
	\taumag\sim \taurad \sim 0.84\ \unitprs\five \text{in the RZ}.
\end{align}

Meanwhile in the CZ, the Reynolds-stress torque has not been measured helioseismically (although could be in the future via ring analysis; e.g., \citealt{Greer2015,Greer2016,Nagashima2020}). Nonetheless, the meridional flow's amplitude $|\upol|\sim10\ {\rm m\ s^{-1}}$ is fairly well known, at least in the upper half of the CZ (e.g., \citealt{Zhao2012,Chen2017,Braun2021}) and so we estimate $\taumc\sim (3/2\pi)\tilde{\rho}|\upol|\rsun\Delta\Omega\cz$, or
\begin{align}\label{eq:taumcest}
	\taurs\sim \taumc\sim \sn{1.2}{6}\ \unitprs\five \text{in the CZ}.
\end{align}
Equations \eqref{eq:tauradest} and \eqref{eq:taumcest} suggest that a diffusively coupled solar CZ and RZ (in which the magnitude of $\taumag$ is similar in both zones) can support the fast magnetic confinement scenario, i.e., $\taurad\sim\taumag\ll \taurs\sim\taumc$. We can further express $\taumag$  from the large-scale non-axisymmetric field in terms of field strength: $\taumag\sim [1/(2\sqrt{2}\pi^2)]|B_{\phi}|^2$. Here, we have (crudely) assumed that $|\vecb_{{\rm pol}}|\sim|B_{\phi}|$, that $r\sin\theta\sim r/\sqrt{2}$, and that the typical length-scale for large-scale field variation is $\sim$$\pi r/2$. Equations \eqref{eq:tauradest} and \eqref{eq:taumcest} then yield
\begin{align}\label{eq:fieldstrength}
	4.8 \gauss\ \lesssim |B_{\phi}| \ll 5800\ \gauss.
\end{align}
Equation \eqref{eq:fieldstrength} states that if the fast magnetic confinement scenario operates in the Sun, we expect the zonal field strength to be significantly less than $5800$ G in the CZ (so as not to disturb the torque balance there) and to diffusively decay to a lower bound of at least $4.8$ G in the tachocline region (to counter radiative spread). The value of the lower bound depends strongly on the actual value of $\Gamma_\odot$ and the value of the upper bound on the reliability of the simulations' prediction that strong field quenches differential rotation.\footnote{If the large-scale solar magnetic field does \textit{not} quench differential rotation when strong enough, we would have no reason to expect $\taumag\ll\taumc$. In fact, the CZ torque balance could be $\taurs+\taumag=0$, in which case $\taumag\sim\taurs$, which would be unconstrained until the Reynolds-stress torque is measured.}

\subsection{Small Skin Depths and Spread of a Permanent Dynamo Field}
Equation \eqref{eq:skindepthmw} shows that, except for $\omega=m\Omega\rz$, any oscillatory component of the solar dynamo has a very small skin depth and thus cannot significantly penetrate into the RZ. This is the reason why prior 1D models like \citet{Barnabe2017} required an $\etaref$ greatly enhanced from its molecular value. Explicitly, we rewrite Equation \eqref{eq:skindepthmw} in dimensional form as
\begin{align}
	\delta_{m\omega}=\left( \frac{2\av{\eta}\rz}{|\omega-m\Omega\rz|}\right)^{1/2}= (0.027\rsun) \pcyc^{1/2},
\end{align}
where here $\pcyc\definealt 2\pi/|\omega-m\Omega\rz|$ and is measured in Gyr. If we require diffusive spread over (say) $\Gamma_\odot=0.05\rsun$, we need $\pcyc\sim1.4$ Gyr. With the solar age at $t_\odot=4.6$ Gyr, such a high $\pcyc$ cannot unambiguously constitute a ``cycle" and instead better corresponds to the permanent component of $\bpol$ (as viewed in the frame rotating with the RZ), here denoted by $\bpolperm$.\footnote{The discussion here implies that the term ``fast" magnetic confinement scenario may be something of an oxymoron; probably ``dynamo" confinement scenario would be a more inclusive term.} There are few, if any, constraints on the solar $|\bpolperm|$, only that it is significantly less than $|\bpol|$ (e.g., \citealt{Usoskin2013}). It is not obvious, however, \textit{how} much less than $|\bpol|$ it really is, and thus whether we can rule out a dynamo confinement scenario entirely if $\etaref$ is not turbulently enhanced. 

For example, even if the solar dynamo were purely axisymmetric and perfectly cyclic with a period of 22 yr, we would expect at most $N_{\rm cyc} = (4.6\ {\rm Gyr})/(22\ {\rm yr})=\sn{2.1}{8}$ cycles since the dynamo turned on. If we assume there have always been random modulations of the cycle amplitude (as are observed throughout recorded history), then we estimate $|\avphi{\bpolperm}|=|\avphi{\bpol}|/\sqrt{N_{\rm{cyc}}} = \sn{6.9}{-5}|\avphi{\bpol}|$. Given Equation \eqref{eq:fieldstrength}, this reduced field strength would only be a factor of $\sim$10 too small to confine the tachocline.\footnote{In other words, from Equation \eqref{eq:fieldstrength}, we compute $4.8/5800=\sn{8.3}{-4}$, which is only $\sim$10 times smaller than $\sn{6.9}{-5}$. This estimate also coheres with \citet{Garaud1999}, who found an amplitude of $|\avphi{\bpol}|\sim0.1$ G in the tachocline region due to ``random-walk" diffusive spread.}

As noted at the end of the last Section \ref{sec:cyclena}, any non-axisymmetric $\bpol$ that co-rotates with the RZ is effectively non-oscillatory and thus contributes to $\bpolperm$. If the fast magnetic confinement scenario is generalized to include non-axisymmetric fields, it thus seems possible that $\bpolperm$ (including all $m$'s) could be significantly larger and more topologically complex than prior estimates like \citet{Garaud1999}. \newtext{It is also worth noting that many prior simulations using a variety of codes (e.g., \citealt{Browning2006,Lawson2015,Beaudoin2018,Bice2020}) have all suggested that large-scale magnetic field accumulates preferentially in the tachocline region. Furthermore, the presence of a tachocline was argued to significantly stabilize the large-scale magnetic fields, sometimes lengthening the dynamo cycle period or even producing time-steady dynamos.}

Active longitudes (preferential solar longitudes at which sunspots emerge; e.g., \citealt{Maunder1905,Svalgaard1975,Bogart1982,Ivanov2007}) are particularly striking as a possible contributor to non-axisymmetric $\bpolperm$. Although it would be a major leap to claim that active longitudes imply a permanent interior partial-wreath structure co-rotating with the RZ (authors have done so nonetheless; e.g., \citealt{Olemskoy2009}), it is intriguing that: (1) they often come in opposite-polarity pairs separated in longitude by $180^\circ$ (e.g., \citealt{Bai2003,Mordinov2004}) and (2) they seem to persist, in a properly chosen rotating frame (or in a frame with time-dependent rotation rate), for long time-scales: 20 years \citep{Henney2002} or even $\sim$100 years \citep{Berdyugina2003}.

Whatever the source of $\bpolperm$, it should penetrate into the RZ much deeper than any skin-depth. Considering the Rayleigh problem (i.e., Stokes' \textit{first} problem, of a boundary plate suddenly jerked from rest), we estimate (for $r\leq r_0$):
\begin{subequations}\label{eq:bpolperm}
\begin{align}
	|\bpolperm|(r) &= |\bpolperm|(r_0) {\rm erfc}\left( \frac{r_0-r}{\delta_{\rm perm}}\right),\\
	\where \delta_{\rm perm} &= \sqrt{4 \av{\eta}\rz t_\odot} = 0.21\rsun.
\end{align}
\end{subequations}
For $r_0-r=\Gamma_\odot=0.05\rsun$, we find ${\rm erfc}(0.05/0.21) = 0.73$, i.e., there should be only a $\sim$27\% reduction in $|\bpolperm|$ over the depth of the tachocline. 

\subsection{Conclusion}
In summary, we have performed a suite of dynamo simulations in which tachocline confinement is achieved if the large-scale non-axisymmetric fields (partial wreaths) are strong enough. These partial-wreath structures cycle with frequencies consistent with advection by a low-latitude retrograde jet. The structures thus appear to cycle from the perspective of the rigidly-rotating RZ and penetrate diffusively downward, with the amplitude of the confining $\bpol$ very well-predicted by the skin-depth Equation \eqref{eq:ampskin}. 

As a whole, the simulations presented here effectively achieve a fast magnetic confinement scenario \citep{ForgcsDajka2001}, which is now generalized to include non-axisymmetric fields and a spread in cycle frequencies. Our work thus offers a significantly wider range of applicability to the fastmagnetic confinement scenario. To further constrain if such a scenario is consistent with observations, we might recommend that future work explore in greater detail the processes giving rise to non-axisymmetric magnetic field (such as active longitudes), and determine observationally how fast active-longitude pairs rotate with respect to the RZ. 

In this discussion section, we have argued that if the magnetic diffusivity is not enhanced, then only an effectively permanent component of the solar dynamo can play a role in tachocline confinement. This component can include both the axisymmetric dynamo field (averaged in time since the birth of the Sun) and, possibly more importantly, any non-axisymmetric field that co-rotates with the RZ. In addition to the fast magnetic confinement scenario, we thus might also recommend exploring a more general (possibly slow) ``dynamo confinement scenario." This would be similar to the model of \citet{Gough1998}, but with the permanent dynamo field taking the place of the primordial field. lt would differ from \citet{Gough1998} mainly in that no primordial field would need to be confined to the RZ.

Finally, in order to make further progress on the numerical side, future simulations need to accomplish several computationally challenging tasks. First, they need to achieve equilibrium that is not diffusively controlled.\footnote{\newtext{ILES simulations, for example those run with the {\eulag} code \citep{Smolarkiewicz2004}, may be the path forward here, since the numerical diffusivities are exceedingly small. Some {\eulag} tachocline simulations [for example, Case MHDs of \citet{Beaudoin2018} and Case RC03 of \citet{Guerrero2016a}] contain large-scale magnetic fields and differential rotation profiles rather similar to the tachocline cases of the present work. This lends some preliminary support to the robustness of our results in the non-diffusively-controlled regime.}} Second, they must be run in the $\sigma\lesssim1$ regime [see Equation \eqref{eq:sigma}]; only then can we assess whether a dynamo confinement scenario can operate in the solar regime of little viscous torque. Finally, simulations must be run with small skin-depths $\delta_{m\omega}$ (i.e., low $\ek$ or high $\prm$). Skin-depths as small as in the Sun would not be possible, but we may at least achieve $\delta_{m\omega}<\Gamma$, which would help confirm if the permanent dynamo field could penetrate deeply enough [possibly according to Equations \eqref{eq:bpolperm}] to confine the tachocline. 
\clearpage
\newpage
\appendix
\restartappendixnumbering
\renewcommand\theHtable{Appendix.\thetable}

\section{Background State}\label{ap:ref}
Note that in this section, we discuss both the non-dimensional and dimensional versions of various quantities. To explicitly distinguish, we denote the dimensional version of a quantity with a ``dim" subscript (quantities like $\cp$ and $\tilde{\nu}$, which are always dimensional, do not require a subscript).  

In terms of the dimensional background state, the perfect-gas law is
\begin{align}\label{eq:idgasdim}
	\prsref\dimm = \left[\frac{(\gamma-1)\cp}{\gamma}\right]\rhoref\dimm\tmpref\dimm,
\end{align}
hydrostatic balance is
\begin{align}\label{eq:hydrdim}
	\frac{d\prsref\dimm}{dr\dimm}=-\rhoref\dimm\gref\dimm,
\end{align}
and the first law of thermodynamics is
\begin{align}\label{eq:firstlawdim}
	\frac{1}{\cp}\left(\frac{dS\dimm}{dr\dimm}\right)=\frac{1}{\gamma}\frac{d\ln\tmpref\dimm}{dr\dimm}-\left(\frac{\gamma-1}{\gamma}\right)\frac{d\ln\rhoref\dimm}{dr\dimm}.
\end{align}

After non-dimensionalizing, Equations \eqref{eq:idgasdim}--\eqref{eq:firstlawdim} take the form
\begin{align}\label{eq:idgas}
	\prsref&= \rhoref\tmpref,
\end{align}
\begin{align}\label{eq:hydr}
	\frac{d\prsref}{dr}=-\di\left(\frac{\gamma}{\gamma-1}\right)\rhoref\,\gref,
\end{align}
and 
\begin{align}\label{eq:firstlaw}
	\dsdr=\frac{1}{\gamma}\frac{d\ln\tmpref}{dr}-\left(\frac{\gamma-1}{\gamma}\right)\frac{d\ln\rhoref}{dr},
\end{align}
where $\sref\definealt\sref\dimm/\cp$ and we recall that $\di\definealt \tilde{g}H/(\cp\tilde{T})$. We combine Equations \eqref{eq:idgas}--\eqref{eq:firstlaw} to yield
\begin{align}
	\frac{d\tmpref}{dr}-\left(\dsdr\right)\tmpref &= -\di \gref,
\end{align}
which has the exact solution [after choosing, without loss of generality, $\sref(r_0)=0$],
\begin{align}\label{eq:tmphat}
	\tmpref &= e^{\sref}\left[\tmpref(r_0) - \di\int_{r_0}^r \gref(x)  e^{-\sref(x)}dx\right].
\end{align}
We then eliminate $\prsref$ from Equations \eqref{eq:idgas}  and \eqref{eq:hydr} to yield
\begin{align}\label{eq:rhohat}
	\rhoref &= \rhoref(r_0) \exp{\left[-\left(\frac{\gamma}{\gamma-1}\right)\sref\right]}\left[\frac{\tmpref\ofr}{\tmpref(r_0)}\right]^{1/(\gamma-1)}. 
\end{align}
There are three equations relating $\rhoref(r_0)$, $\tmpref(r_0)$, $\di$, $\gamma$, $\beta$, and $N_\rho$: two from our  choice of non-dimensionalization---$(4\pi/V\cz)\int_{r_0}^{r\out}\rhoref(r)r^2dr=1$ and $(4\pi/V\cz)\int_{r_0}^{r\out}\tmpref(r)r^2dr=1$, where $V\cz\definealt (4\pi/3)(r\out^3-r_0^3)=(4\pi/3)[(1-\beta^3)/(1-\beta)^3]$---and one from the definition $N_\rho\definealt\ln[\rhoref(r_0)/\rhoref(r\out)]$. Thus, $\rhoref(r_0)$, $\tmpref(r_0)$, and $\di$ may be regarded as functions of $\gamma$, $\beta$, and $N_\rho$. For the values given in Table \ref{tab:inputnondim} (and our choices for $\gref$ and $\dsdrline$ given below), we explicitly find $\rhoref(r_0)=2.67$, $\tmpref(r_0)=2.04$, and $\di=1.72$.

For $\gref\ofr\propto1/r^2$ and the condition $(4\pi/V\cz)\int_{r_0}^{r\out}\gref(r)r^2dr=1$, we require
\begin{align}
	\gref(r) = \left[ \frac{1-\beta^3}{3(1-\beta)^3}    \right]\frac{1}{r^2}.
\end{align}
To model the transition from convective stability to instability at the base of the CZ, we choose $\dsdrline$ to be zero in the CZ, a constant positive (near-unity) value in the RZ, and continuously matched in between:
\begin{equation}
	\frac{d\sref}{dr} = \begin{cases}
		\Sigma & r\leq r_0 - \delta\\
		\Sigma\bigg{\{}1 - \left(1 - \Big{(}\frac{r-r_0}{\delta}\Big{)}^2\right]^2\bigg{\}} & r_0 - \delta < r < r_0\\
		0 & r\geq r_0,
	\end{cases}
	\label{eq:dsdr}
\end{equation}
where $\Sigma=0.453$ (note that $\Sigma$ is not really a free parameter, since it can always be absorbed into the fluid control parameter $\rm Bu$). The choice of quartic matching ensures that the ultimate stability transition (determined by the total entropy gradient $\dsdrline+d\avaltsph{\spert}/dr$ in the equilibrated state) is never too far from $r_0$. By contrast, for a tanh matching [$\dsdrline=(\Sigma/2)(1-\tanh{[(r-r_0)/\delta]})$; e.g., \citealt{Korre2021}], the stability transition can occur significantly above $r_0$. In our cases, it could occur as high up as $r_0+5\delta$, since $d\avsph{S}/dr$ is generally $10^4$--$10^5$ times smaller than $\dsdrline$ and $(1/2)[1-\tanh{(5)}]\approx\sn{5}{-5}$. 

\begin{table}
	\caption{Derived non-dimensional parameters and time-scales for our simulations, which can be obtained from Table \ref{tab:inputnondim} and the form of the reference state. These include the Taylor number $\ta$, the Rayleigh number $\raf$, and the convective Rossby number $\roc$. In the lower part of the table, all time-scales are non-dimensional (i.e., scaled by $\Omega_0^{-1}$). The diffusion times ($\pnu$, $\peta$, etc.) estimate the time for different diffusive processes across different sub-domains (CZ, RZ, or full-shell) of the simulation.}
	\label{tab:inputnondimderiv}
	\centering
	\begin{tabular}{l  l l}
		\hline
		Parameter & Definition & Value\\
		\hline
		$r\inn$ & $(2\beta-1)/(1-\beta)$ & 2.15\\
		$r_0$ & $\beta/(1-\beta)$ & 3.15\\
		$r\out$ & $1/(1-\beta)$ & 4.15\\
		$N_{\rho,{\rm{RZ}}}$ & $\ln[\rhoref(r\inn)/\rhoref(r_0)]$ & 2.08\\
		$\di$ & $\tilde{g}{H}/\cp\tilde{T}$ & 1.72 \\		
		$\ta$ & $\ek^{-2}$ & $\sn{8.80}{5}$\\
		$\raf$ & $\rafmod\pr/\ek^2$ & $\sn{5.62}{5}$\\
		$\roc$ & $\sqrt{\rafmod}/2$ & 0.400\\
		$\ek\rz$ & $ \av{\nuref}\rz\ek$ & $\sn{3.47}{-4}$\\
		$\sigma$ & $\sqrt{\rm \pr B}/2$ & 79.6 \\
		\hline
		$\prot$ & rotation period & $2\pi$\\
		$\pnu=\pkappa$ & $(4/\av{\nuref}_{\rm full})/\ek$  & $779\ \prot$ \\
		$\peta$ & $(4/\av{\etaref}_{\rm full})\prm/\ek$  & $779$ to $6240\ \prot$ \\			
		$\pnu_{\rm,CZ}=\pkappa_{\rm,CZ}$ & $1/\ek$  & $149\ \prot$  \\
		$\peta_{\rm,CZ}$ & $\prm/\ek$  & $149$ to $1190\ \prot$  \\
		$\pnu_{\rm,RZ}=\pkappa_{\rm,RZ}$ & $1/\ek\rz$  & $459\ \prot$  \\
		$\peta_{\rm,RZ}$ & $\prm/\ek\rz$  & $459$ to $3670\ \prot$  \\
		$\pes$ & $\pkappa_{\rm,RZ} \rm Bu/4$ & $\sn{2.91}{6}\ \prot$  \\ 
		\hline
	\end{tabular}
\end{table}

With $\dsdrline$ and $\gref$ chosen, we numerically integrate Equations \eqref{eq:tmphat} and \eqref{eq:rhohat} to find $\rhoref$ and $\tmpref$. This approach to defining the background state (also used by \citealt{Korre2021}) has the main advantage that hydrostatic balance is satisfied everywhere, even in the transition region. This stands in contrast to polytropic matching (e.g., \citealt{Racine2011,Guerrero2016a}).

Note that equation \eqref{eq:dsdr} also defines the buoyancy frequency through 
\begin{align}
	\frac{\nsqref}{\gref} = \frac{\dsdrline}{\langle\gref\dsdrline\rangle\rz},
\end{align}
where $\langle\gref\dsdrline\rangle\rz=0.597$. 

We choose $\qref(r)$ to occupy primarily the CZ:
\begin{equation}
	\qref= \frac{c}{2}\left[1 + \tanh{\left(\frac{r-r_0}{\delta_{\rm{heat}}}\right)}\rhoref\tmpref\right],
	\label{eq:heat}
\end{equation}
where $c=0.944$. This value of $c$ is required because $\qref=Q\dimm H/\widetilde{F_{\rm{nr}}}$. The definition $(\fluxnr)\dimm\equiv(1/Hr^2)\int_r^{r\out}Q\dimm(x)x^2dx$ then yields $1/c = (2\pi/V\cz)\int_{r_0}^{r\out}(1/r^2)\int_r^{r\out}f(x)x^2dxdr$, where $f(x)\definealt 1+\tanh[(x-r_0)/\delta_{\rm{heat}}]$. 

Table \ref{tab:inputnondimderiv} gives some additional (derivative) input parameters that can be computed from the parameters of Table \ref{tab:inputnondim} and the form of the reference state just described. 

 \begin{figure*}
	\centering
	\includegraphics[width=7.25in]{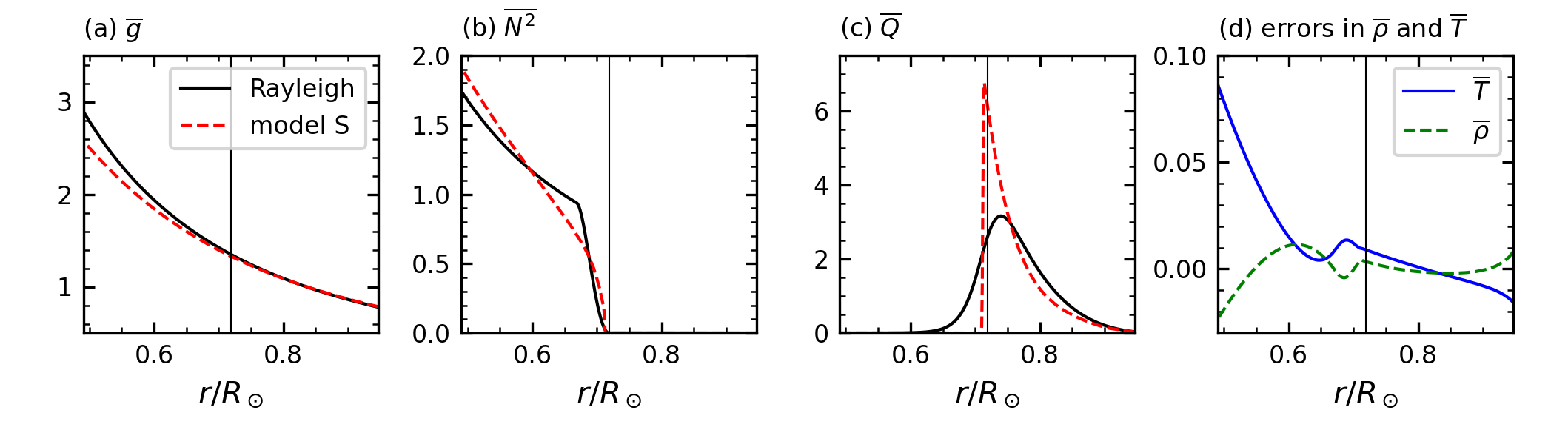}
	\caption{(a)--(c) Non-dimensional reference state (solid black curves) compared to Model S (dashed red curves). (d) Relative errors (compared to Model S) in our reference-state for $\rhoref\ofr$ and $\tmpref\ofr$, with the error defined as, e.g., $(\rhoref - \rhoref_S)/\rhoref_S$. In all panels, the vertical line denotes the CZ--RZ interface $r=r_0$.}
	\label{fig:model_S}
\end{figure*}
We compare this reference state to the standard solar Model S \citep{ChristensenDalsgaard1996} in Figure \ref{fig:model_S}. Note that Model S profiles (denoted by an ``S" subscript) are originally in dimensional form.

For the dimensional molecular diffusivities associated with Model S, we define
\begin{subequations}
\begin{align}
	\nuref_S &\definealt \sn{1.2}{-16}\ \frac{(\tmpref_S/\kelv)^{5/2}}{\rhoref_S/(\unitrho)}\ \stoke,\\
	\kapparef_S &\definealt \frac{16\sigma_{\rm SB}\tmpref_S^3}{3\overline{\chi}_S\rhoref_S^2(\cp)_S},\\
	\etaref_S &\definealt \sn{5.2}{11}\ \frac{\ln\Lambda}{(\tmpref_S/\kelv)^{3/2}}\ \stoke,
\end{align}
\end{subequations}
where $\sigma_{SB}\definealt\sn{5.67}{-5}\ \rm{erg\ cm^{-2}\ s^{-1}\ K^{-4}}$ is the Stefan-Boltzmann constant and $\overline{\chi}_S$ is the opacity from Model S. The forms of the molecular viscosity $\nuref_S$ and the radiative thermal diffusivity $\kappa_S$ are given in (e.g.) \citet{Parker1979} via \citet{Miesch2005}. The form of $\etaref_S$ is given in (e.g.) \citet{Spitzer1962}. The Coulomb logarithm $\ln\Lambda$ is tabulated by (e.g.) \citet{Stix2002}, and we approximate $\ln\Lambda\approx2.5 + r/r_0$ (see \citealt{Garaud1999}). 

To non-dimensionalize Model S, we take $(R_\odot)\dimm=\sn{6.96}{10}$ cm and set $(r\inn)\dimm=0.491(\rsun)\dimm$, $(r_0)\dimm=0.719(\rsun)\dimm$, and $(r\out)\dimm=0.947(\rsun)\dimm$ [and thus $H=0.228(\rsun)\dimm$]. This choice means we compare to the bottom three density scale-heights of Model S's CZ, i.e., $\ln{\{ \rhoref_S[(r_0)\dimm]/\rhoref_S[(r\out)\dimm]\}}=3$. For a given reference-state quantity $\psi$, we then define $\av{\psi}\cz$ (or $\tilde{\psi}$) as a volume average of $\psi$ over $((r_0)\dimm, (r\out)\dimm)$ and $\av{\psi}\rz$ as a volume average of $\psi$ over $[(r\inn)\dimm, (r_0)\dimm]$. We then scale $\rhoref_S$, $\tmpref_S$, and $\gref_S$ by $\tilde{\rho}_S$, $\tilde{T}_S$, and $\tilde{g}_S$ (respectively), $\qref_S$ by $(\fluxnrtilde)_S/H$, and $(\overline{N^2})_S$ by $\av{(N^2)_S}\rz$. Then for the rest of this section, the Model S profiles denote their non-dimensional forms.

Figure \ref{fig:model_S} shows that {\rayleigh}'s non-dimensional reference state is fairly solar-like, being equivalent to the adiabatic polytrope from our prior work (e.g., \citealt{Featherstone2016a, Orvedahl2018, Matilsky2019, Hindman2020}). The biggest discrepancies occur near $r=r_0$, where our reference state has relatively wide and smooth transitions in $\nsqref$ and $\qref$ compared to the narrow and sharp transitions from Model S. 

Note that the background diffusivities from the simulations' reference state, $\nuref$, $\kapparef$, and $\etaref$, are specified independently from the thermodynamic profiles. We choose all simulation diffusivities to increase with radius like $\rhoref^{-1/2}$ (and of course they are normalized to have a volume-average over the CZ of 1). Note that this choice does not in any sense correspond to the non-dimensional Model S profiles, $\nuref_S$, $\kapparef_S$, and $\etaref_S$.

\section{Dimensional Solar Analog}\label{ap:analog}
Here, we ``re-dimensionalize" the models considered in the current paper to match the presentation of cases H and 4.00 in \citetalias{Matilsky2022}. A non-dimensional simulation can be re-dimensionalized by assuming dimensional values for quantities like $H$ and $\Omega_0$ and then computing the associated scales for the fluid variables (e.g., $[\spert]$ and $[\vecu]$) and reference-state profiles (e.g., $\tilde{\rho}$ and $\tilde{T}$), as described in Section \ref{sec:exp}. To list the input dimensional quantities in the conventional way, we define the luminosity $L\definealt\tilde{Q}V\dimm$, where $V\dimm\definealt(4\pi H^3/3)(r\out^3-r\inn^3)$. $L$ is typically the control parameter that sets $\widetilde{F_{\rm{nr}}}$. We also define the stellar mass $M\definealt[(1-\beta^3)/3(1-\beta)^3](H^2/G)\tilde{g}$, so $\gref\dimm=GM/H^2r^2$. $M$ is typically the parameter that sets $\tilde{g}$. The full set of dimensional input parameters is then: $H$, $\Omega_0$, $L$, $M$, $\cp$, $\tilde{\rho}$, $\tilde{T}$, $\tilde{\nu}$, $\tilde{\kappa}$, $\tilde{\eta}$, and $\langle N^2\rangle\rz$. 

Some of the input dimensional parameters are obviously redundant, so there are infinitely many ways to re-dimensionalize. The only requirement is that the chosen dimensional values be consistent with the input non-dimensional numbers. Historically, we in the solar and stellar communities have chosen stellar-like dimensional values for as many parameters as possible except for the diffusivities, which are chosen to be unrealistically high. This choice is exemplified in Table \ref{tab:solaranalog}, which contains the scaling employed in \citetalias{Matilsky2022}. Most of the chosen parameters are solar-like, except for the diffusivities. The rotation rate $\Omega_0$ is chosen to be about three times higher than the solar Carrington value. 

The inherent non-uniqueness associated with dimensional simulations is one of the main reasons we report only the non-dimensional versions of the simulations in this work. For example, comparing the simulated $\vecb\dimm$ (measured in G) to an observed $\vecb$ at the solar surface (also measured in G) is fundamentally ambiguous. For, we could have re-dimensionalized using the values in Table \ref{tab:solaranalog}, but instead chosen $\Omega_0\rightarrow \Omega_\odot$, $L\rightarrow\lsun/27$, $\langle N^2\rangle\rz\rightarrow \langle N^2\rangle\rz/3$, $\tilde{\nu}\rightarrow\tilde{\nu}/3$, $\tilde{\kappa}\rightarrow\tilde{\kappa}/3$, and $\tilde{\eta}\rightarrow\tilde{\eta}/3$. This would have yielded dynamically identical simulations, but with all values of $\vecb\dimm$ three times smaller.

	\begin{table}
	\caption{\citetalias{Matilsky2022}'s re-dimensionalization of our models. Recall $[S] = \Delta S \definealt\fluxnrtilde H/\tilde{\rho}\tilde{T}\tilde{\kappa}$, $[P] = \tilde{\rho}(2\Omega_0H)^2$, $[\vecu] = \Omega_0H$, $[\vecb] = \sqrt{\mu \tilde{\rho}}(\Omega_0H)$, and $\mu=4\pi$ in Gaussian units.}
	\label{tab:solaranalog}
	\centering
		\begin{tabular}{l l l}
			Quantity & Model S value & Dimensional analog value \\
			\hline
			 $H$&      $\sn{1.59}{10}\ \cm$        & $\sn{1.59}{10}\ \cm$  \\
			 $\Omega_0$ &     $\sn{2.70}{-6}\  \rm rad\ s^{-1}$       &  $\sn{8.61}{-6}\  \rm rad\ s^{-1}$\\
			 $\prot$ &     26.9 days       &  8.45 days  \\
			 $L$ &       $\sn{3.40}{33}\ \rm erg\ s^{-1}$    &  $\lsun\definealt\sn{3.85}{33}\ \rm erg\ s^{-1}$\\
			 $\fluxnrtilde$ &    $\sn{7.12}{10}\ \rm erg\ \cm^{-2}\ s^{-1}$        &  $\sn{6.79}{10}\ \rm erg\ \cm^{-2}\ s^{-1}$\\
			 $M$ &      $\sn{1.97}{33}$ g      &  $\msun\definealt\sn{1.99}{33}$ g\\
			 $\tilde{g}$ &       $\sn{3.90}{4}\ \rm \cm\ \sec^{-2}$      &  $\sn{3.93}{4}\ \rm \cm\ \sec^{-2}$\\ 
			 $\tilde{\rho}$ &     $\sn{6.79}{-2}\ \unitrho$       &  $\sn{6.75}{-2}\ \unitrho$\\ 
			 $\av{\rho}\rz$ &     $0.523\ \unitrho$       &  $0.520\ \unitrho$\\ 
			$\tilde{T}$ &       $\sn{1.06}{6}\ \kelv$     &  $\sn{1.03}{6}\ \kelv$\\ 			
			$\cp$ &       $\sn{3.54}{8}\ \unitent$      &  $\sn{3.50}{8}\ \unitent$ \\
			 $\av{N^2}\rz$ &     $\sn{2.03}{-6}\ \rm (rad\ s^{-1})^2$       &  $\sn{1.88}{-6}\ \rm (rad\ s^{-1})^2$\\[2pt]
			 $\tilde{\nu}$ &      $2.21\ \stoke$      &  $\sn{2.31}{12}\ \stoke$\\ 
			 $\tilde{\kappa}$ &       $\sn{3.90}{6}\ \stoke$      &  $\sn{2.31}{12}\ \stoke$\\ 
			 $\tilde{\eta}$ &      $\sn{3.09}{3}\ \stoke$       & (0.289--2.31)$\sn{}{12}\ \stoke$\\
			 $\av{\nu}\rz$ &      $4.15\ \stoke$      &  $\sn{7.51}{11}\ \stoke$\\ 
			 $\av{\kappa}\rz$ &      $\sn{9.70}{6}\ \stoke$      &  $\sn{7.51}{11}\ \stoke$\\ 
			 $\av{\eta}\rz$ &      $\sn{3.61}{2}\ \stoke$      &  (0.939--7.51)$\sn{}{11}\ \stoke$\\ 
			 \hline
			 $(\pnu_{\rm,RZ})\dimm$ & $\sn{1.92}{12}$ years & 10.6 years\\
			 $(\peta_{\rm,RZ})\dimm$ & $\sn{2.21}{10}$ years & (10.6--84.9) years\\
			 $(\pes)\dimm$ & $\sn{5.72}{10}$ years & $\sn{6.73}{4}$ years\\			 
			\hline
			$[\spert]$ &       $\sn{4.04}{9}\ \unitent$     &  $\sn{6.69}{3}\ \unitent$\\
			$[\prspert]$ &      $\sn{1.24}{8}\ \uniten$      &  $\sn{2.01}{10}\ \uniten$\\
			$[\vecu]$&        $\sn{4.28}{4}\ \rm{m\ s^{-1}}$     &  $\sn{1.40}{3}\ \rm{m\ s^{-1}}$\\
			$[\vecb]$ &      $\sn{3.95}{4}\ \gauss$       &  $\sn{1.26}{5}\ \gauss$\\
			\hline
		\end{tabular}
	\end{table}

\section{Output Non-Dimensional Numbers}\label{ap:outputnd}
We define the Reynolds (Re), Rossby (Ro), and magnetic Reynolds ($\rem$) numbers, separately for the mean and fluctuating flows:
\begin{align}
	\re_{\rm{mean}} \definealt \frac{(\avphi{\vecu})\rms}{\ek},           \five             \re_{\rm{fluc}}    \definealt \frac{(\vecu^\prime)\rms}{\ek},\label{eq:re}\\
	\ro_{\rm{mean}}\definealt \frac{(\avphi{\vecom})\rms}{2},      \five             \ro_{\rm{fluc}}      \definealt \frac{(\vecom^\prime)\rms}{2},\label{eq:ro}\\
	\rem_{,\rm{mean}}\definealt\re_{\rm{mean}}\prm,   \five        \rem_{,\rm{fluc}}   \definealt \re_{\rm{fluc}}\prm\label{eq:rem},
\end{align}
where $\vecom\definealt\curl\vecu$ is the vorticity and the mean in the rms is taken in volume (over the CZ or RZ) and in time over the equilibrated state. Table \ref{tab:outputnondim2} contains the values of these numbers for each simulation considered in this work. 

\begin{table*}
	\caption{Output non-dimensional numbers, defined in Equations \eqref{eq:re}--\eqref{eq:rem}, for all simulations. The number values in the CZ and RZ are given separately.}
	\label{tab:outputnondim2}
	\centering
	\begin{tabular}{*{14}{l}  }
		\hline

Case   &   H   &   1.00   &   1.05   &   1.06   &   1.08   &   1.33   &   1.67   &   2.00   &   3.00   &   4.00   &   6.00   &   8.00  \\
regime   &   -   &   W   &   W   &   W   &   M    &   M   &   M   &   M   &   S   &   S   &   S   &   S  \\
\hline
\multicolumn{13}{c}{CZ non-dimensional numbers}\\
${\re}_{\rm mean}$   &   173.9   &   174.1   &   173.8   &   173.1   &   85.56   &   71.06   &   66.36   &   65.93   &   45.51   &   40.63   &   34.69   &   30.28  \\
${\re}_{\rm fluc}$   &   68.74   &   68.76   &   68.47   &   68.58   &   58.05   &   57.81   &   57.65   &   57.57   &   57.25   &   56.77   &   56.33   &   55.95  \\
${\ro}_{\rm mean}$   &   0.144   &   0.145   &   0.144   &   0.144   &   0.097   &   0.085   &   0.081   &   0.079   &   0.063   &   0.057   &   0.050   &   0.045  \\
${\ro}_{\rm fluc}$   &   0.438   &   0.438   &   0.437   &   0.438   &   0.436   &   0.435   &   0.433   &   0.432   &   0.428   &   0.423   &   0.423   &   0.423  \\
${\rem}_{,\rm mean}$   &   -   &   174.1   &   183.2   &   184.3   &   92.09   &   94.75   &   110.6   &   131.9   &   136.5   &   162.5   &   208.2   &   242.2  \\
${\rem}_{,\rm fluc}$   &   -   &   68.76   &   72.15   &   73.03   &   62.49   &   77.08   &   96.09   &   115.1   &   171.8   &   227.1   &   338.0   &   447.6  \\
\hline
\multicolumn{13}{c}{RZ non-dimensional numbers}\\
${\re}_{\rm mean}$   &   219.1   &   219.8   &   219.1   &   216.9   &   61.86   &   57.08   &   51.85   &   53.40   &   13.42   &   10.84   &   10.06   &   8.757  \\
${\re}_{\rm fluc}$   &   45.19   &   45.15   &   44.48   &   44.86   &   16.15   &   13.05   &   12.26   &   11.97   &   10.43   &   10.21   &   9.850   &   9.534  \\
${\ro}_{\rm mean}$   &   0.070   &   0.070   &   0.070   &   0.069   &   0.025   &   0.021   &   0.020   &   0.020   &   9.3e-3   &   8.0e-3   &   7.1e-3   &   6.4e-3  \\
${\ro}_{\rm fluc}$   &   0.076   &   0.076   &   0.075   &   0.075   &   0.043   &   0.040   &   0.039   &   0.039   &   0.036   &   0.035   &   0.035   &   0.034  \\
${\rem}_{,\rm mean}$   &   -   &   219.8   &   230.9   &   231.0   &   66.58   &   76.11   &   86.42   &   106.8   &   40.27   &   43.35   &   60.37   &   70.06  \\
${\rem}_{,\rm fluc}$   &   -   &   45.15   &   46.88   &   47.78   &   17.39   &   17.40   &   20.43   &   23.95   &   31.30   &   40.83   &   59.10   &   76.28  \\

		\hline
	\end{tabular}
\end{table*}

\clearpage
\newpage

\acknowledgments
We thank the following (undoubtedly incomplete) list of individuals for helpful discussions of the solar tachocline:  Pascale Garaud, Mark Miesch, Lydia Korre, Nicholas Featherstone, Gustavo Guerrero, Connor Bice,  Sacha Brun, Antoine Strugarek, Catherine Blume, Matthew Browning, Steven Tobias, David Hughes, and J{\o}rgen Christensen-Dalsgaard. This work was partly done in collaboration with the COFFIES DRIVE Science Center (NASA grant 80NSSC22M0162). L.I.M. and N.H.B. thank the Isaac Newton Institute for Mathematical Sciences (Cambridge, UK) for support and hospitality during the 2022 Programme ``Frontiers in Dynamo Theory: from the Earth to the Sun and Stars", where part of the work on this paper was undertaken. L.I.M. was primarily supported during this work by a National Science Foundation Astronomy \& Astrophysics Postdoctoral Fellowship under award AST-2202253, as well as a Future Investigators in NASA Earth and Space Sciences Technology (FINESST) award 80NSSC19K1428. The computations integral to this work were supported by NASA grant 80NSSC18K1127. This research was further supported by NASA grants 80NSSC18K1125, 80NSSC19K0267, and 80NSSC20K0193. Computational resources were provided by the NASA High-End Computing (HEC) Program through the NASA Advanced Supercomputing (NAS) Division at Ames Research Center. {\rayleigh} is hosted and receives support from the Computational Infrastructure for Geodynamics (CIG), which is supported by the National Science Foundation awards NSF-0949446, NSF-1550901, and NSF-2149126. The input files, final checkpoints, and some data-analysis products (averaged data and basic plots) for all simulations are publicly accessible via Zenodo \citep{Matilsky2023c}.


	
\end{document}